\documentclass{elsart}
\usepackage[dvips]{graphicx}
\usepackage{natbib}
\usepackage{color}
\usepackage{amssymb}
\usepackage{amsmath}
\usepackage{epsf}
\usepackage{subfigure}
\begin{document}

\begin{frontmatter}
\title{Scaling with system size of the Lyapunov exponents for the Hamiltonian Mean Field model}
\author{Thanos Manos$^1$} and \author{Stefano Ruffo$^{1,2,3}$}

\address{1. Dipartimento di Energetica ``S. Stecco", Universit{\`a} di Firenze,\\ via S.~Marta
3, 50139, Firenze, Italy}

\address{2. INFN, Sezione di Firenze, Italy}

\address{3. Centro Interdipartimentale per lo Studio delle Dinamiche Complesse (CSDC),
Universit{\`a} di Firenze, Firenze, Italy}

\thanks[thanos]{E-mail: thanosm@master.math.upatras.gr}
\thanks[stefano]{E-mail: stefano.ruffo@unifi.it}

%=============================================
\begin{abstract}

The Hamiltonian Mean Field (HMF) model is a prototype for systems
with long-range interactions. It describes the motion of $N$
particles moving on a ring, coupled through an infinite-range
potential. The model has a second order phase transition at the
energy density $U_c=3/4$ and its dynamics is exactly described by the Vlasov
equation in the $N \to \infty$ limit. Its chaotic properties have
been investigated in the past, but the determination of the scaling
with $N$ of the Lyapunov Spectrum (LS) of the model remains a
challenging open problem. We here show that the $N^{-1/3}$ scaling
of the Maximal Lyapunov Exponent (MLE), found in previous numerical
and analytical studies, extends to the full LS; not only, scaling is
``precocious" for the LS, meaning that it becomes manifest for a
much smaller number of particles than the one needed to check the
scaling for the MLE. Besides that, the $N^{-1/3}$ scaling appears to
be valid not only for $U>U_c$, as suggested by theoretical
approaches based on a random matrix approximation, but also below a
threshold energy $U_t \approx 0.2$. Using a recently proposed method
(GALI) devised to rapidly check the chaotic or regular nature of an
orbit, we find that $U_t$ is also the energy at which a sharp
transition from {\it weak} to {\it strong} chaos is present in the
phase-space of the model. Around this energy the phase of the vector
order parameter of the model becomes strongly time dependent,
inducing a significant untrapping of particles from a nonlinear
resonance.

\end{abstract}

\begin{keyword}
Hamiltonian Mean Field (HMF) model, Lyapunov spectra, GALI method
\end{keyword}

\end{frontmatter}

%=============================================
\section{Introduction}
\label{intro}

The dynamical and statistical properties of systems with long-range
interactions have recently attracted considerable attention (see
\citet{CDRrev:2009} for a recent review). The Hamiltonian Mean Field
(HMF) model \citep{Ant_Ruf:1995} is considered as a prototype for
some of the observed dynamical effects. The model describes the
motion of $N$ fully coupled particles lying on a ring. The system
presents a second order phase transition at the critical energy density
$U_c=E_c/N=3/4$ from a low-energy phase, where the particles are
clustered, to a high-energy gaseous phase where the particles are
uniformly distributed on the ring. Several different effects have
been studied for this model, which have been then found to be
generic for a large class of systems with long-range interactions,
including self-gravitating systems, unscreened plasmas,
two-dimensional hydrodynamics, etc..

The question on which we concentrate in this paper is the behavior
of the Lyapunov exponents. This question has been investigated for a
long time, already in some of the first papers on the HMF model.
However, a general understanding of the chaotic properties of this
model remains a challenge. Since we will be mostly concerned with
the $N \to \infty$ behavior of the Lyapunov exponents, this question
has a relevance also for the Vlasov equation, which is known to
describe exactly the dynamics of the HMF model in this limit
\citep{CDRrev:2009}.

Let us briefly summarize the state of the art knowledge on the
Lyapunov exponents for the HMF model. The Maximal Lyapunov Exponent
(MLE) increases with energy up to the critical energy density $U_c$
and then decreases for larger energies \citep{Yamaguchi:96}. In the
whole high energy phase it drops to zero in the large $N$ limit as
$N^{-1/3}$ \citep{Latoraetal:1998}. An analytical estimate of the
MLE, which takes into account microcanonical averages of suitable
geometrical observables \citep{Pettini:2007}, was proposed by
\citet{Firpo:1998}. A preliminary study of the scaling properties of
the Lyapunov Spectra (LS) and of the Kolmogorov-Sinai (KS) entropy
was performed by \citet{Latoraetal:1999}. Lyapunov
instability of a modified HMF model, the so-called $\alpha$-HMF, was
first studied by \citet{Ante_Tsa:1998} and further analyzed by
\citet{Campa1:2001,Campa2:2002}. The $N^{-1/3}$ scaling has been
shown to derive from a random matrix approximation
\citep{Firpo:2001,Anteneodo1:2001,Anteneodo2:2002, Anteneodo3:2003}.
A supersymmetric approach \citep{Tanase:2003} suggests that the MLE
might vanish in the $N \to \infty$ for all energies. A tendency of
the model towards integrability when $N$ increases, meaning a
vanishing MLE, has been recently emphasized using a Poincar\'e
return map approach \citep{Bache_etal:2008}. Lyapunov exponents and
the corresponding eigenmodes of the HMF model have been also
recently studied in the Vlasov $N \to \infty$ limit
\citep{Paska_etal:2009}.

The aim of this paper is to present a careful assessment of the scaling properties with $N$
of the MLE, the LS and the KS entropy of the HMF model in various energy ranges and in conditions
as close as possible to Boltzmann-Gibbs thermal equilibrium. With respect to previous studies, we have put
more effort in establishing the equilibrium state, knowing that this is not easily reached
due to the presence of quasi-stationary states \citep{CDRrev:2009}. These latter are
out--of--equilibrium states in which the system remains trapped for a time that increases with $N$,
slowing down indefinitely the relaxation to thermal equilibrium.

An alternative approach, designed for fast detection of chaotic orbits, has been recently
proposed \citep{GALI:2007}. The method, which has been called Generalized Alignment Index (GALI),
allows one to distinguish between regular and chaotic motion using short-time numerical
integrations. The GALI method takes into account more than one deviation vector
and measures the time evolution of the volume of the parallelepiped whose edges are these
normalized vectors. In this paper we use the GALI method to detect the fraction of chaotic
orbits on a constant energy surface of the HMF model.

The paper is organized as follows. In Section~\ref{HMF_model} we briefly introduce
the HMF model and we discuss the main features of the Boltzmann-Gibbs statistical equilibrium state. In
Section~\ref{Lyap_exp} we recall how the LS is computed, in order to make the paper self-consistent.
Afterwards, in Section~\ref{LS_KSres}, we present the core results of this paper, by
discussing the scaling with $N$ of the MLE, the LS and the KS entropy. In Section~\ref{GALI_def}
we describe the GALI chaos detection method and we test its efficiency
by considering the HMF model with a small number of particles.  In Section~\ref{GALI_res} we use
the GALI method for the estimation of the fraction of chaotic orbits in the model's phase space.
Our conclusions are finally summarized in Section~\ref{concl}.

\section{The HMF model and its statistical equilibrium}
\label{HMF_model}

The Hamiltonian Mean Field (HMF) model \citep{Ant_Ruf:1995} describes a system of point masses
moving on a ring and interacting through an infinite-range potential. The Hamiltonian of the
model is
\begin{equation}
\label{HMF}
H = \sum_{i=1}^{N} \frac{p_i^2}{2}
+\frac{1}{2N}\sum_{i,j=1}^{N}[1-\cos(\theta_i-\theta_j)],
\end{equation}
where $\theta_i \in (\pi,\pi]$ is the coordinate of the $i$-th particle
and $p_i$ its conjugate momentum. The system has only two constants
of the motion: total energy $H$ and total momentum $\sum_i p_i$ (in
the following we will always choose a zero total momentum without
loosing generality). As the force vanishes when $i=j$, the particles
do not collide but smoothly cross each other. The {\it order
parameter} of this model is the ``magnetization" vector
\begin{equation}\label{magn}
\mathbf{M}=\frac{1}{N} \sum_{i=1}^{N} (\cos \theta_i,\sin \theta_i)=(M_x,M_y)~,
\end{equation}
and, after defining its phase $\phi = \arctan M_y / M_x$ and modulus $M=\sqrt{M_x^2+M_y^2}$,
the equations of motion can be written as
\begin{eqnarray}\label{HME:eq_motion}
   \left\{
    \begin{array}{ll}
     \dot{\theta}_i = p_i \\
     \dot{p}_i = -M_x \sin \theta_i  + M_y \cos \theta_i= -M \sin (\theta_i - \phi)~.
    \end{array}
   \right.
\end{eqnarray}

We solve numerically the equations of motion of the HMF model using an optimized
fourth-order symplectic integrator \citep{MacLa_Ate:1992} with time step $h=0.05$, which
typically gives a relative energy fluctuation $\Delta E/E \sim 10^{-4}$.

The equilibrium statistical mechanics of the HMF model has been
thoroughly studied (see \citet{CDRrev:2009} for a review). The model
has a second order phase transition at the critical energy density $U_c=3/4$ from a
low-energy clustered phase where $M \neq 0$ to a high-energy gaseous
phase where $M=0$. Being this phase transition of second order, it cannot be associated
with a negative specific heat. The canonical and microcanonical ensemble give equivalent 
results \citep{CDRrev:2009}.

In the mean-field limit, $N \to \infty$, the dynamics of the HMF model is
fully described by the single particle distribution function $f(\theta, p, t)$,
where $(\theta,p)$ are the canonically conjugate Eulerian coordinates of the single
particle phase space. The single particle distribution function obeys a Vlasov equation.
Among many stationary solutions of the Vlasov equation, a particular one is the
Boltzmann-Gibbs equilibrium distribution
\begin{equation}
\label{equilibrium}
f_{eq} (\theta,p) = \frac{\sqrt{\beta}}{(2 \pi)^{3/2} I_0(\beta M)} \exp [\beta (\frac{p^2}{2} - M \cos
(\theta - \phi)],
\end{equation}
where $\beta=1/T$ is the inverse temperature and $I_0$ is the modified Bessel
function of order zero. This distribution is also stable with $M=0$ for $\beta \leq \beta_c=2$ and
with $M \neq 0$ for $\beta > \beta_c$. Therefore, $\beta_c$ can be interpreted
as the inverse critical temperature of the second order phase transition,
corresponding to the critical energy $U_c$.

In order to reach the equilibrium distribution (\ref{equilibrium}) on a reasonably short
time scale, we have initialized positions with a Gaussian distribution and we
have computed the corresponding potential energy. We then determined the
appropriate width of the Gaussian distribution of momenta which
gave the energy we wanted to achieve. This initial state was then evolved for a
long time in order for the equilibrium distribution of
Eq.~(\ref{equilibrium}) to be reproduced with a sufficient precision.
For instance, for the $N=100$ particle case, we can reproduce the equilibrium
temperature with a relative error of a few percent, which is compatible
with the statistical error, expected to be of the order $N^{-1/2}$.

\section{The Lyapunov spectrum}
\label{Lyap_exp}

For the sake of completeness, we briefly recall how the Lyapunov spectrum
of a $2N$-dimensional flow is defined and computed \citep{Ben:1980a,Ben:1980b}, with
reference to the HMF model.

The $2N$-dimensional phase-space coordinate is
\begin{equation}
\mathbf{x} \equiv \left(\theta_1,\dots,\theta_N,p_1,\dots,p_N \right)
\end{equation}
and the phase-space flow is generated by the system of autonomous first-order
differential equations (\ref{HME:eq_motion}), which, using $\mathbf{x}$, can
be written as
\begin{equation}
\label{dif_eq}
\frac{d\mathbf{x}(t)}{dt}=F(\mathbf{x}(t))~,
\end{equation}
where $F$ is the velocity field of the flow.

The evolution equations for the deviation vector
\begin{equation}
\mathbf{w} \equiv \left(\delta\theta_1,\dots,\delta\theta_N,\delta p_1,\dots,\delta p_N \right)
\end{equation}
are
\begin{equation}
\label{var_eq}
\frac{d\mathbf{w}}{dt}=\tilde{J}(\mathbf{x}(t))\mathbf{w}~,
\end{equation}
where the $2N \times 2N$ Jacobian matrix of the flow $\tilde{J}(\mathbf{x}(t))=\partial F / \partial \mathbf{x}$
is given by
\begin{equation}
\tilde{J}= \left(
  \begin{array}{cc}
    0 & I \\
    J & 0 \\
  \end{array}
  \right)~,
\end{equation}
with the $N \times N$ Jacobian $J_{ij}=-\partial^2 V / \partial \theta_i \partial \theta_j$ given by
\begin{eqnarray}
\label{eq_motion}
   \left\{
    \begin{array}{ll}
     J_{ii} = -\cos (\theta_i)M_x - \sin (\theta_i)M_y + \frac{1}{N}\\
     J_{ij} = \frac{1}{N} \cos(\theta_i- \theta_j),\quad \text{if} \quad
     i \neq j~,
    \end{array}
   \right.
\end{eqnarray}
and $I$ the $N \times N$ identity matrix.

The linear evolution equations (\ref{var_eq}) are integrated with initial value $\mathbf{w}(0)$,
which physically represents the ``small" initial difference between two nearby orbits of Eq.~(\ref{dif_eq}).
Numerically, this integration is performed in parallel with that of the orbit $\mathbf{x}(t)$,
since $\tilde{J}$ depends on it. One typically uses the same integration algorithm.

The Maximal Lyapunov Exponent (MLE) is then defined as
\begin{equation}
\label{LE}
\lambda (\mathbf {x}(0))=\lim_{t \rightarrow \infty} \frac{1}{t}
\ln \frac{\|\mathbf{w}(t)\|}{\|\mathbf{w}(0)\|}.
\end{equation}
Practically, since $\|\mathbf{w}(t)\|$ diverges exponentially, one
has to renormalize it at regular time intervals. The computation of the MLE is
performed by averaging the renormalization factors.

Furthermore, when varying $\|\mathbf{w}(0)\|$, one can obtain at most $2N$
Lyapunov exponents, which can be ordered in size
\begin{equation}
\lambda_{1} \geq \lambda_{2} ... \geq \lambda_{2N}~,
\end{equation}
and constitute the so-called Lyapunov Spectrum (LS), with $\lambda_{1}=\lambda (\mathbf {x}(0))$.

For Hamiltonian flows, the LS is endowed of the following symmetry property
\begin{equation}
\label{lce_sym}
\lambda_{i}=-\lambda_{2N-i+1}, \quad i=1,2,...,2N~.
\end{equation}

It is then enough to compute the first $N$ exponents. Among them, for the HMF model,
two are zero because of energy and momentum conservation. For a chaotic orbit all
the others are typically positive.

However, because of the numerical instability which tends to align the deviation
vector $\mathbf{w}(t)$ along the most expanding direction, one always obtains $\lambda_1$
and has not access to the full LS.

The trick to perform numerically the calculation of the full LS was found by
\citet{Ben:1980a,Ben:1980b} and amounts to compute the hypervolume
\begin{equation}
\label{volumeV}
\mathcal{V}_{p}(t)=\mathbf{w}_{1}(t)\wedge \mathbf{w}_{2}(t)\wedge...\wedge
\mathbf{w}_{p}(t),
\end{equation}
for $p$ deviation vectors $\mathbf{w}_{1}(t),...,\mathbf{w}_{p}(t)$.

Then, the sum of the first $p$ Lyapunov exponents
\begin{equation}
\lambda^{(p)}=\lambda_{1}+\lambda_{2}+...+\lambda_{p}
\end{equation}
is given by
\begin{equation}
\label{Lyap_eq}
\lambda^{(p)} (\mathbf {x}(0))=\lim_{t \rightarrow \infty}\frac{1}{t} \ln
\frac{\|\mathcal{V}_{p}(t)\|}{\|\mathcal{V}_{p}(0)\|}~.
\end{equation}

A recent review about the Lyapunov exponents and their calculation can be found in \citet{Sko:LE}.

Once the LS is known, the Kolmogorov-Sinai (KS) entropy, which is the
rate of information production, is given by
\begin{equation}
\label{SKS}
S_{KS}= \sum_{i=1}^N \lambda_i~.
\end{equation}

\section{Scaling with $N$ of the MLE, the LS and the KS entropy}
\label{LS_KSres}

As mentioned in the Introduction, it has been proposed that in the
whole high energy region $U \ge U_c$ the MLE should vanish as $N^{-1/3}$.
We want to show here that this scaling law is obeyed also by the
full LS.

\begin{figure*}[h]
\centering
  \includegraphics[width=7.cm]{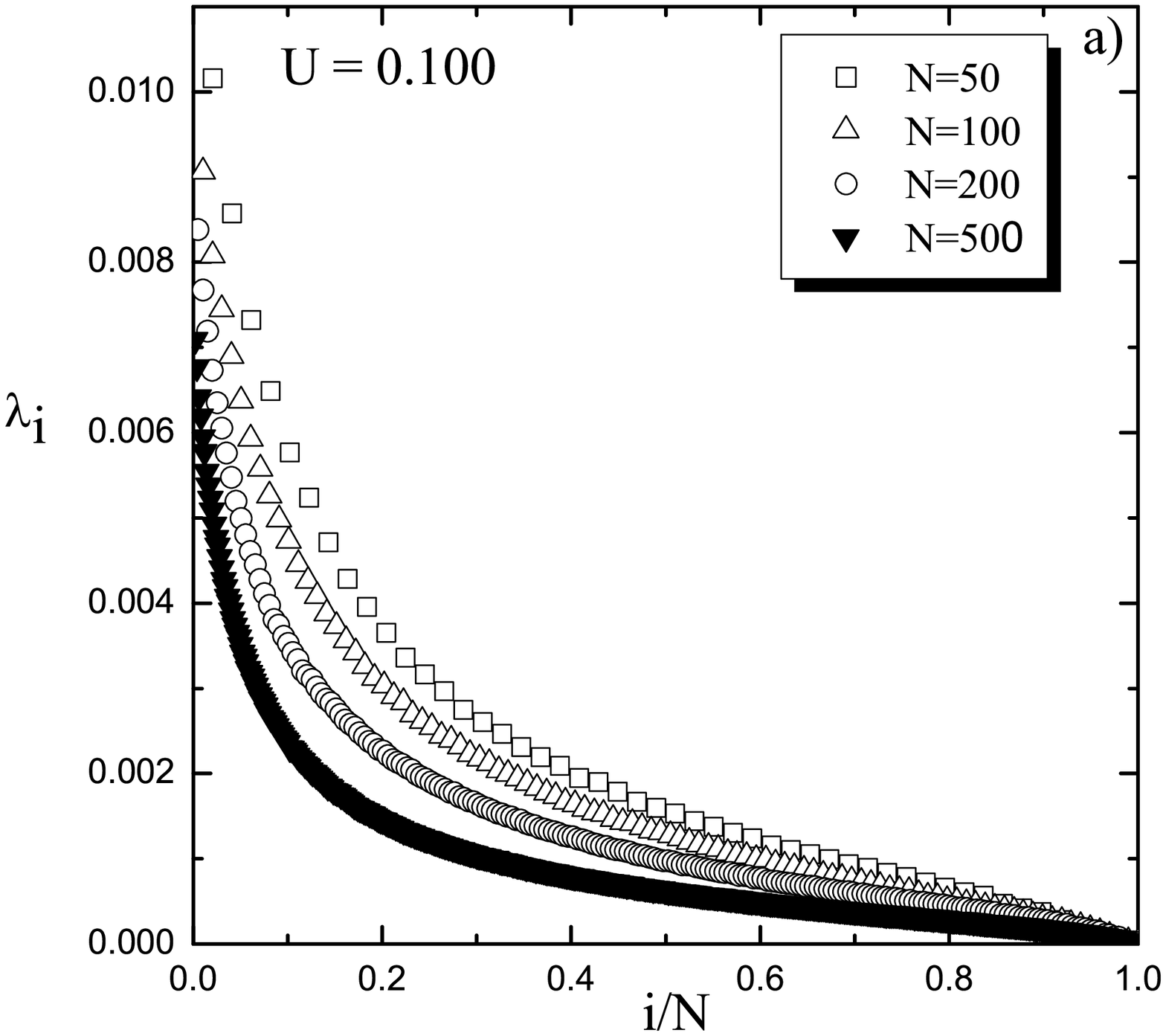}\hspace{-1.75cm}
  \includegraphics[width=7.cm]{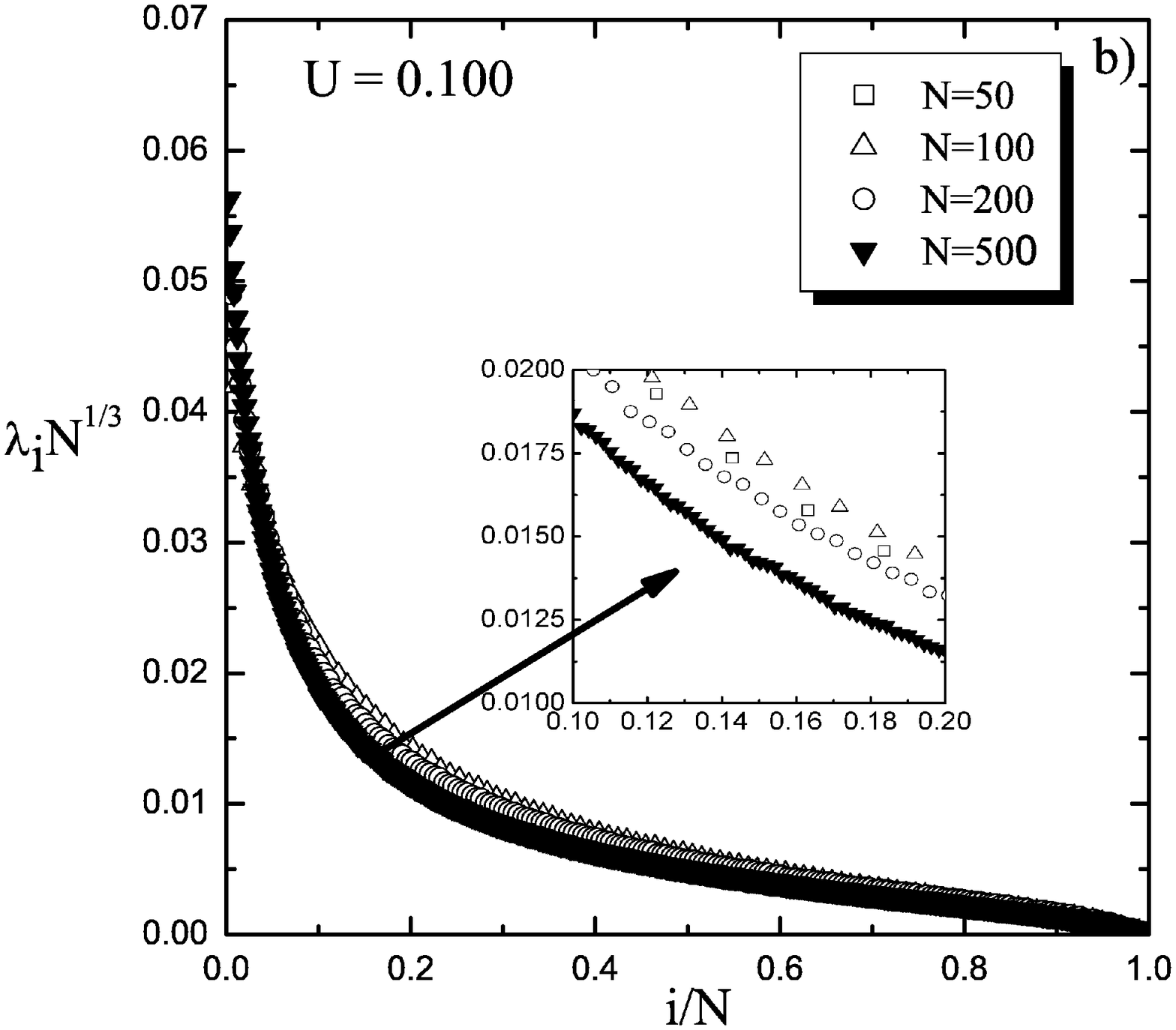}\\
  \includegraphics[width=7.cm]{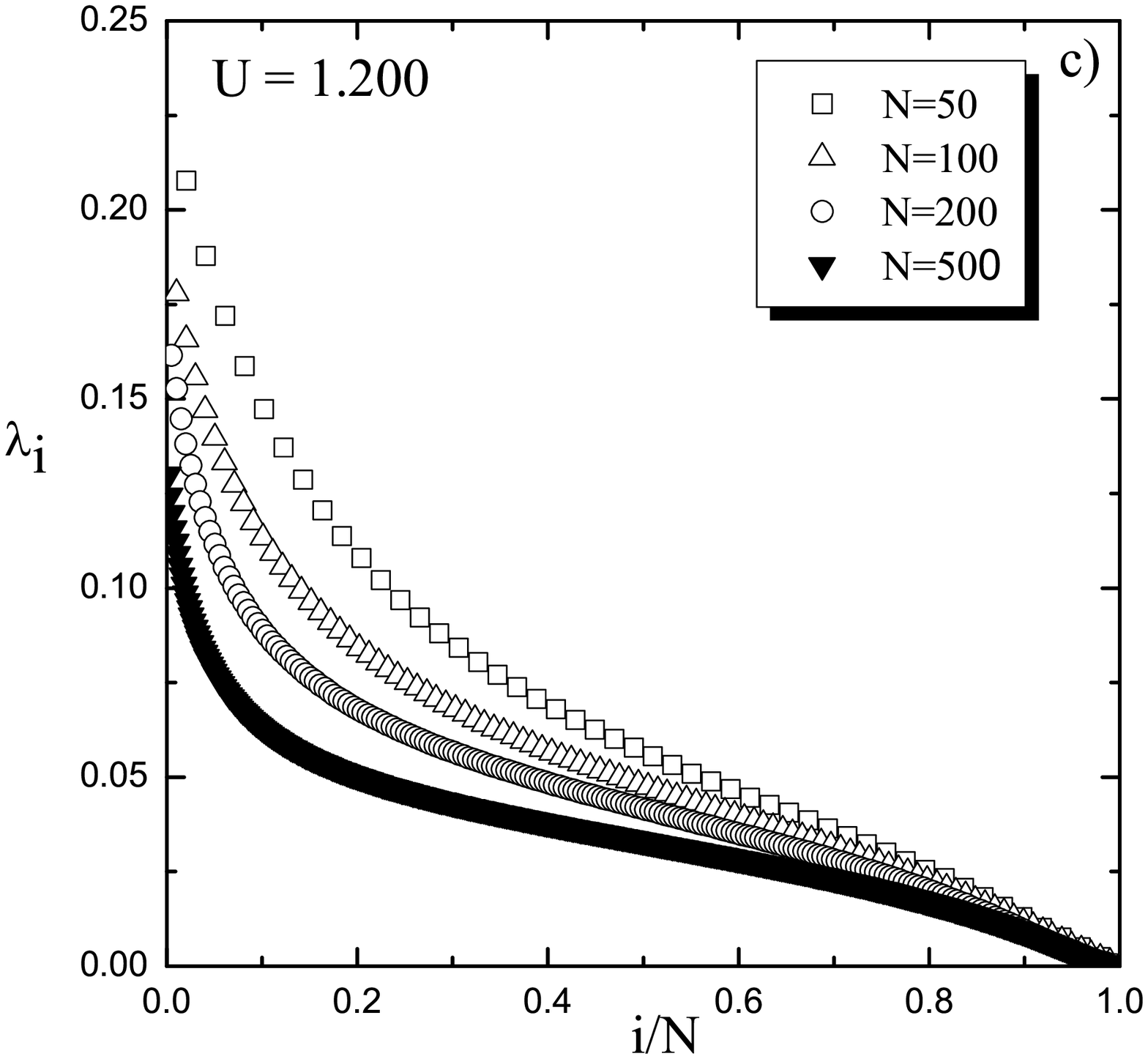}\hspace{-1.75cm}
  \includegraphics[width=7.cm]{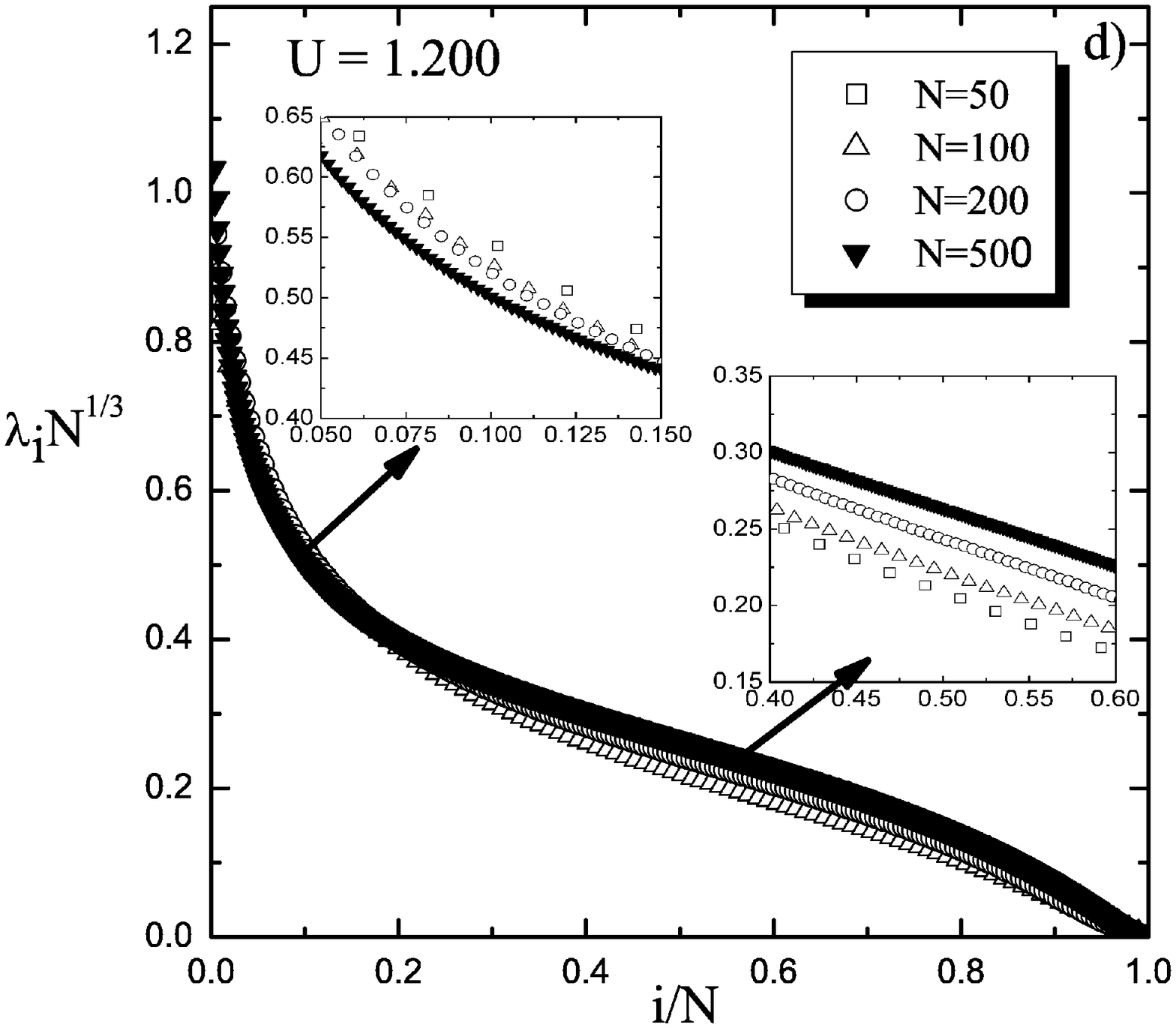}
  \caption{LS vs. $i/N$ for $U=0.1$ (panel a) and $U=1.2$ (panel c).
  For the same energies we also show the LS multiplied by $N^{1/3}$
  (panels b and d). The insets show details of the size dependence with more
  resolution.}
   \label{LE_sev}
\end{figure*}

In panels a) and c) of Fig.~\ref{LE_sev} we plot $\lambda_i$
vs. $i/N$, increasing $N$ from $N=50$ to $N=500$, for $U=0.1$ (panel a)
and $U=1.2$ (panel c).
The total integration time was $0.5\times 10^6$ with time
step $h=0.05$, which provides a good convergence for the LS and a good
accuracy for the energy conservation as well. The relative error
in the determination of the MLE is of the order of $1 \%$.
We observe a general decreasing trend, showing
a significant size dependence of the LS. This size dependence is
almost completely eliminated when multiplying the LS by $N^{1/3}$, as shown in panels b)
and d) of the same figure. We observe a reasonably good ``data collapse" in the
upper and lower parts of the spectrum. However, the insets in panels 
b) and d) show that the size dependence is not yet completely eliminated and
remains of the order of $10 \%$.
The insets reveal another interesting
feature: while the convergence to an asymptotic LS is from above for the full LS at $U=0.1$, 
at the energy $U=1.2$ it is from above only for the largest exponents and it is 
instead from below for the smaller exponents.

%=============================================
%=============================================

\begin{figure*}[h]
\centering
\includegraphics[width=7.cm]{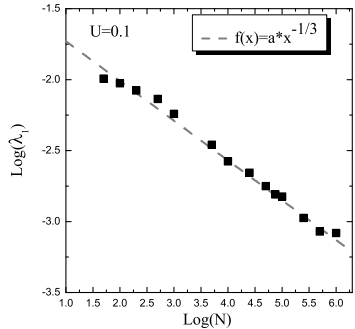}
  \caption{Logarithm of the MLE vs. $\log N$ for $U=0.1$, showing the $N^{-1/3}$ scaling. Both
  logarithms are in base $10$.}
\label{max_01}
\end{figure*}

In Fig.~\ref{max_01} we show the $N$ dependence of $\lambda_1$ for
$U=0.1$, which is well fitted by the scaling law $N^{-1/3}$,
confirming the result obtained by looking at the full LS. However,
one has to reach much larger system sizes (of the order of $10^6$)
in order to check the scaling law, while the convergence to the
asymptotic LS as $N$ increases is observed for much smaller system
sizes (up to $N=500$).

Other numerical experiments
\citep{Latoraetal:1998,Ante_Tsa:1998,Anteneodo3:2003} show that this scaling
law is present also for energies above $U_c$, but the quality of the data
is never comparable with the one obtained here for $U=0.1$.

While the scaling in the high energy range can be justified using a random
matrix approximation \citep{Latoraetal:1998,Ante_Tsa:1998,Anteneodo1:2001} and a theory using geometric properties
of the phase-space \citep{Pettini:2007,Firpo:1998}, no theoretical approach exists for the
explanation of the $N^{-1/3}$ scaling at low energies. On the other hand, the
only theoretical approach valid in this energy region \citep{Firpo:1998} would
predict a strictly positive MLE in the $N \to \infty$ limit.

%=============================================
%=============================================

\begin{figure*}[h!]
%\centering
  \includegraphics[width=6.cm]{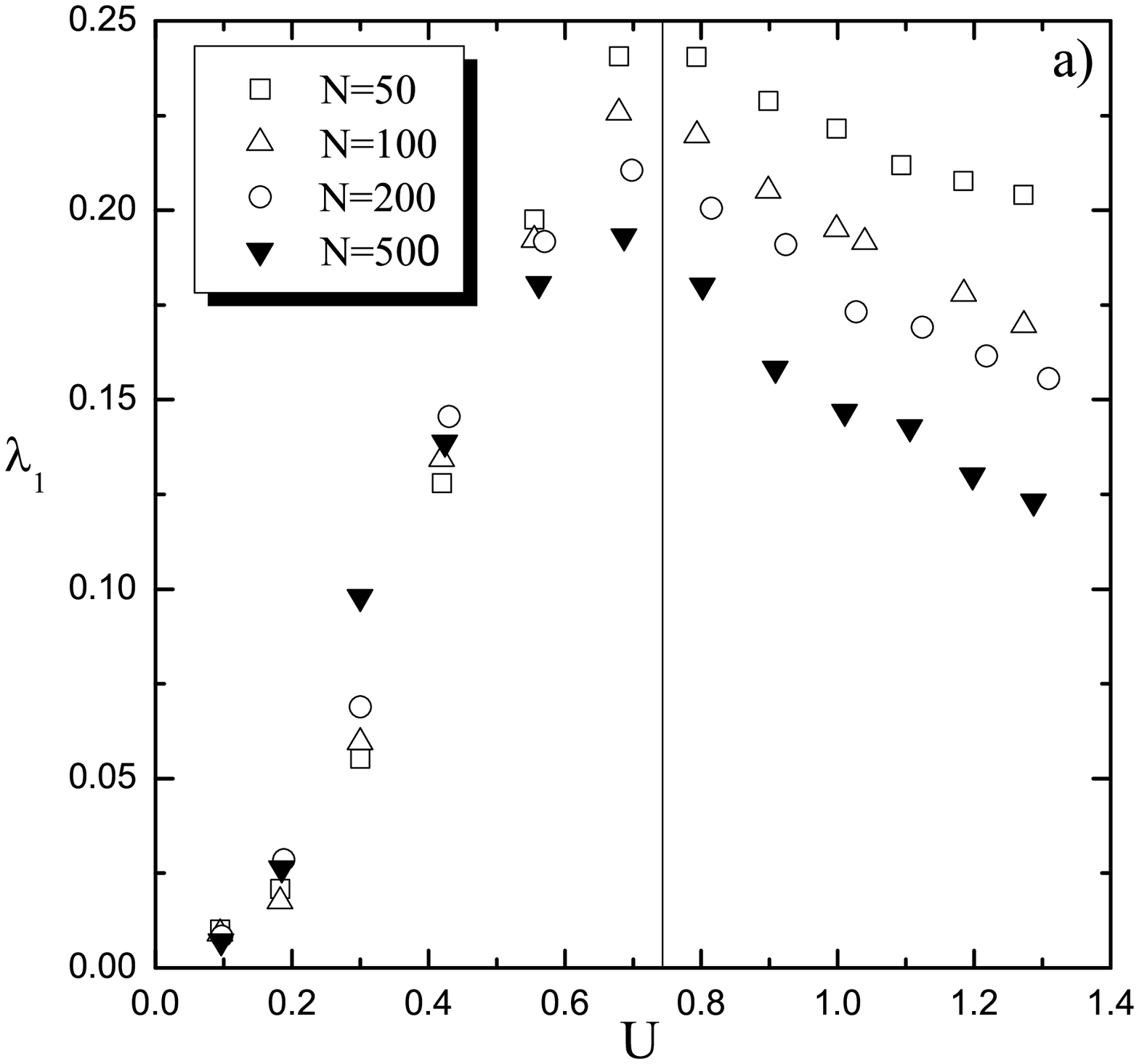}\hspace{-1.75cm}
  \includegraphics[width=6.cm]{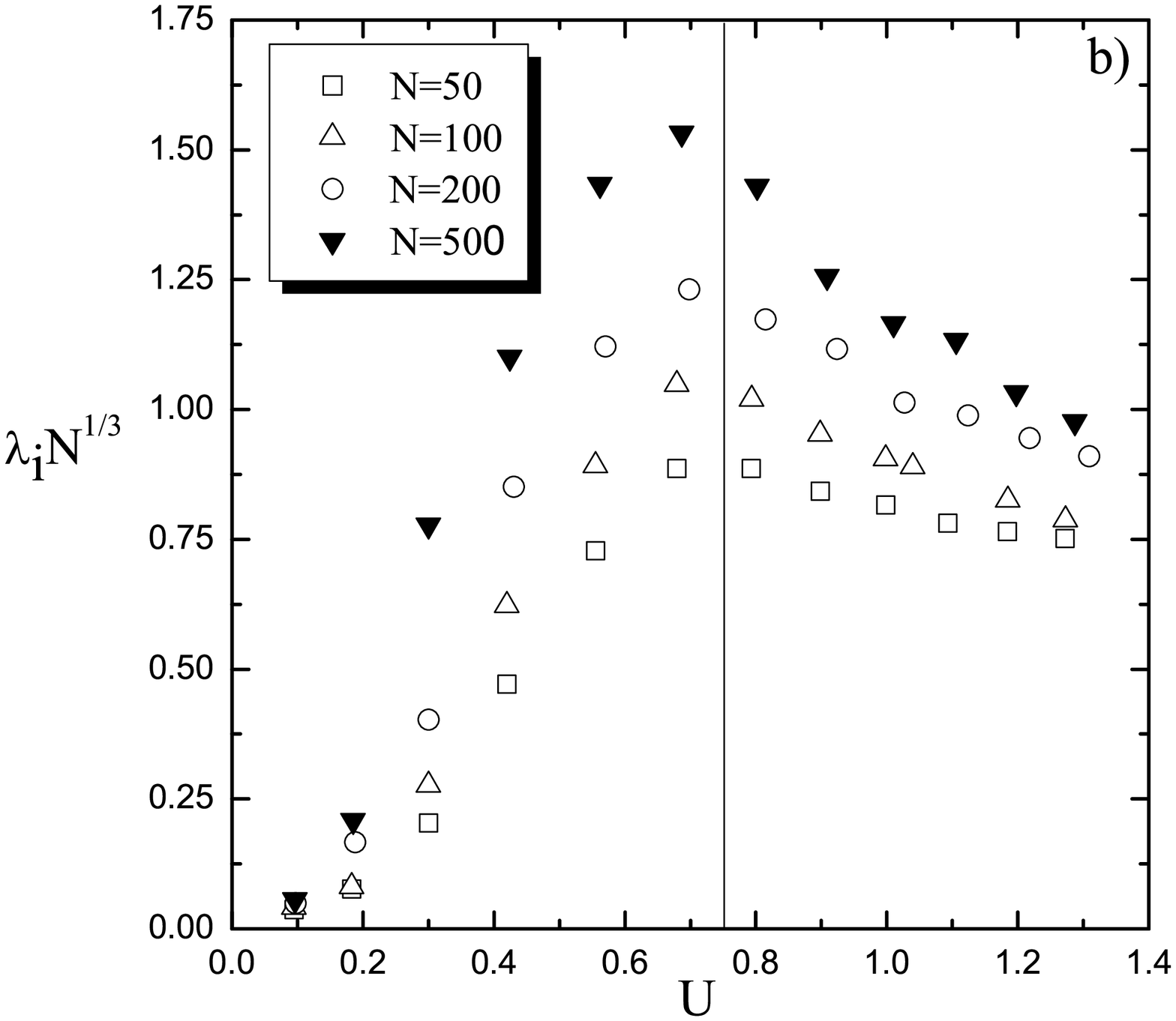}\hspace{-1.75cm}
  \includegraphics[width=6.cm]{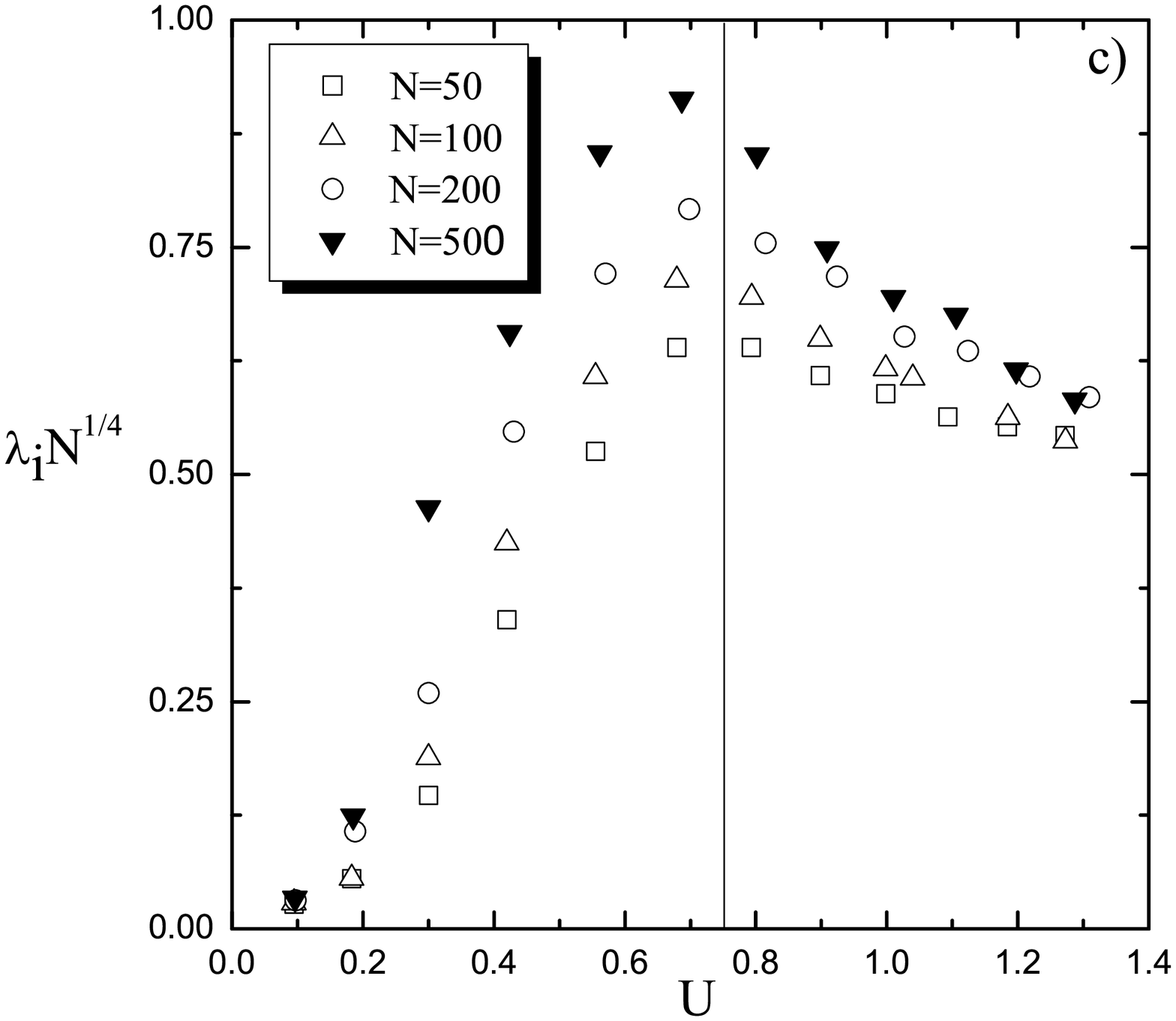}
  \caption{a) The MLE, $\lambda_1$, as a function
  of $U$ for various system sizes $N=50, 100, 200, 500$ (the vertical
  line shows the critical energy $U_c=3/4$).
  b) MLE multiplied by the factor $N^{1/3}$ vs. $U$.
  c) MLE multiplied by the factor $N^{1/4}$ vs. $U$.}
   \label{LE_ab}
\end{figure*}

In Fig.~\ref{LE_ab}a), we show the MLE as a function
of $U$ for various system sizes: $N=50, 100, 200, 500$. One
observes a sharp increase of the MLE around the energy
$U_t \approx 0.2$. We'll come back to comment about this
energy value in Section~\ref{GALI_res}. The MLE is peaked
around the critical energy $U_c$ and shows, in this range
of sizes, a weak size dependence for $U<U_c$ while for $U>U_c$
it systematically and monotonically decreases with system size.
In panel b) of the same figure, we check if the $N^{-1/3}$ scaling
found for the LS works also for the MLE, by multiplying all MLE
by the factor $N^{1/3}$. It turns out that this scaling does not
work well for this relatively small number of particles. An ``indermediate"
scaling with a factor $N^{-1/4}$ looks slightly better in the
high energy range, see Fig.~\ref{LE_ab}c.

As a partial conclusion, we could state that the MLE shows a
size dependence over the whole energy range, but the $N^{-1/3}$
scaling does not emerge easily from the data. This is at variance
with the analysis above of the LS, for which this scaling was more
evident. This is why we speak of ``precocious" scaling of the
LS, meaning that the scaling is obtained for moderately large
system sizes, while the scaling of the MLE, as shown e.g. in Fig.~\ref{max_01},
needs much larger systems sizes to be numerically emphasized.

%=============================================
%=============================================

\begin{figure*}[h!]
  \includegraphics[width=6.cm]{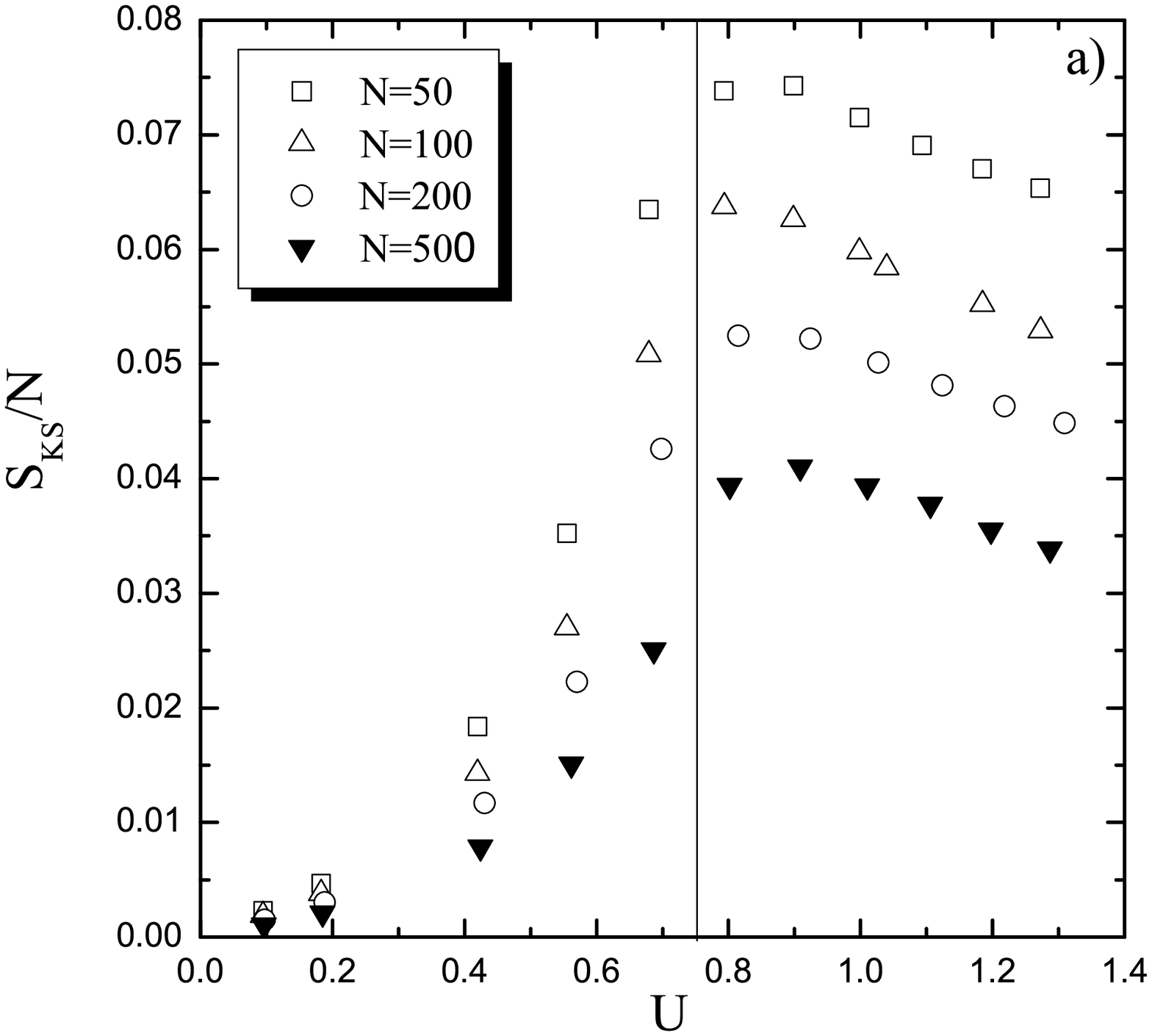}\hspace{-1.75cm}
  \includegraphics[width=6.cm]{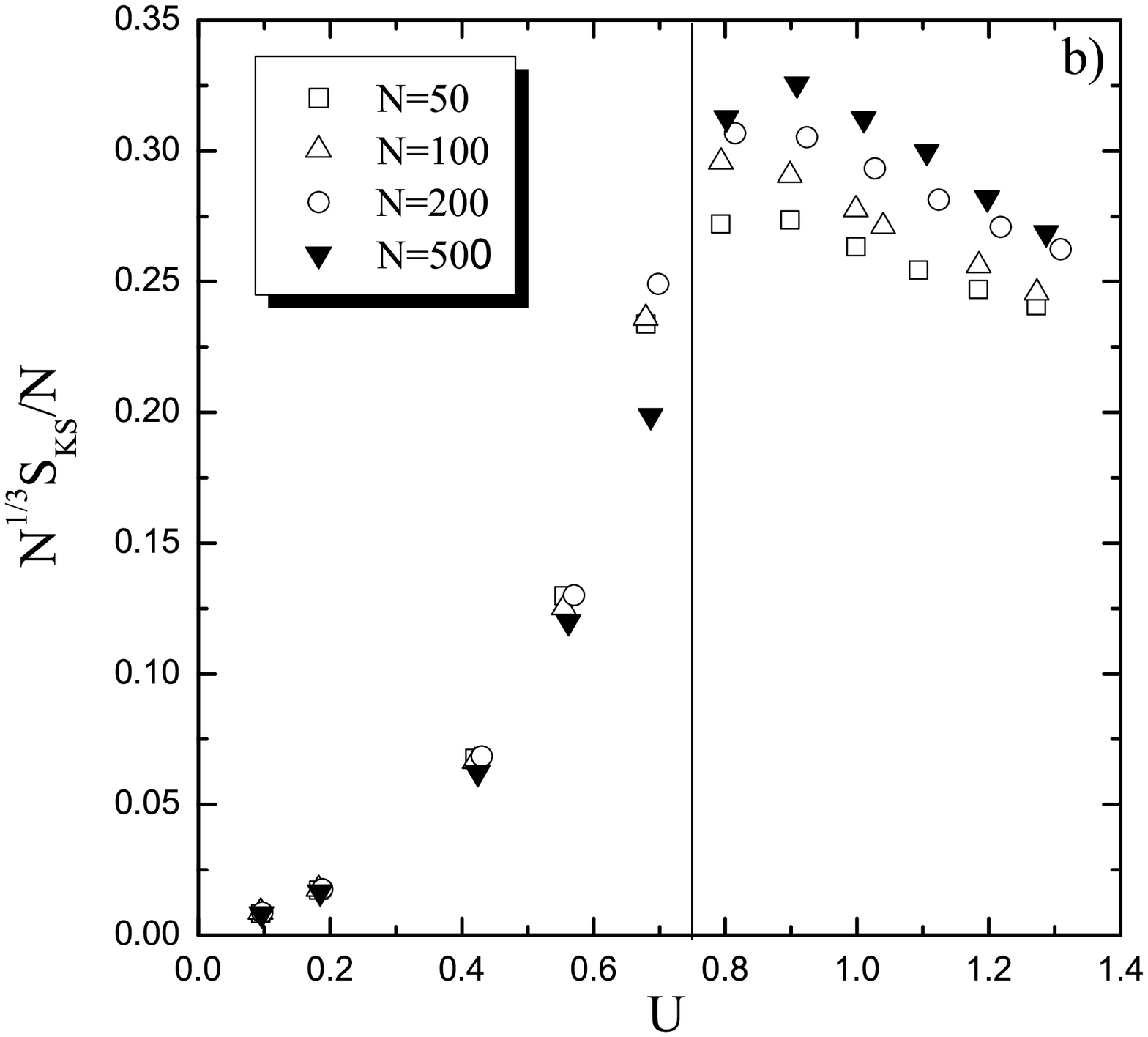}\hspace{-1.75cm}
  \includegraphics[width=6.cm]{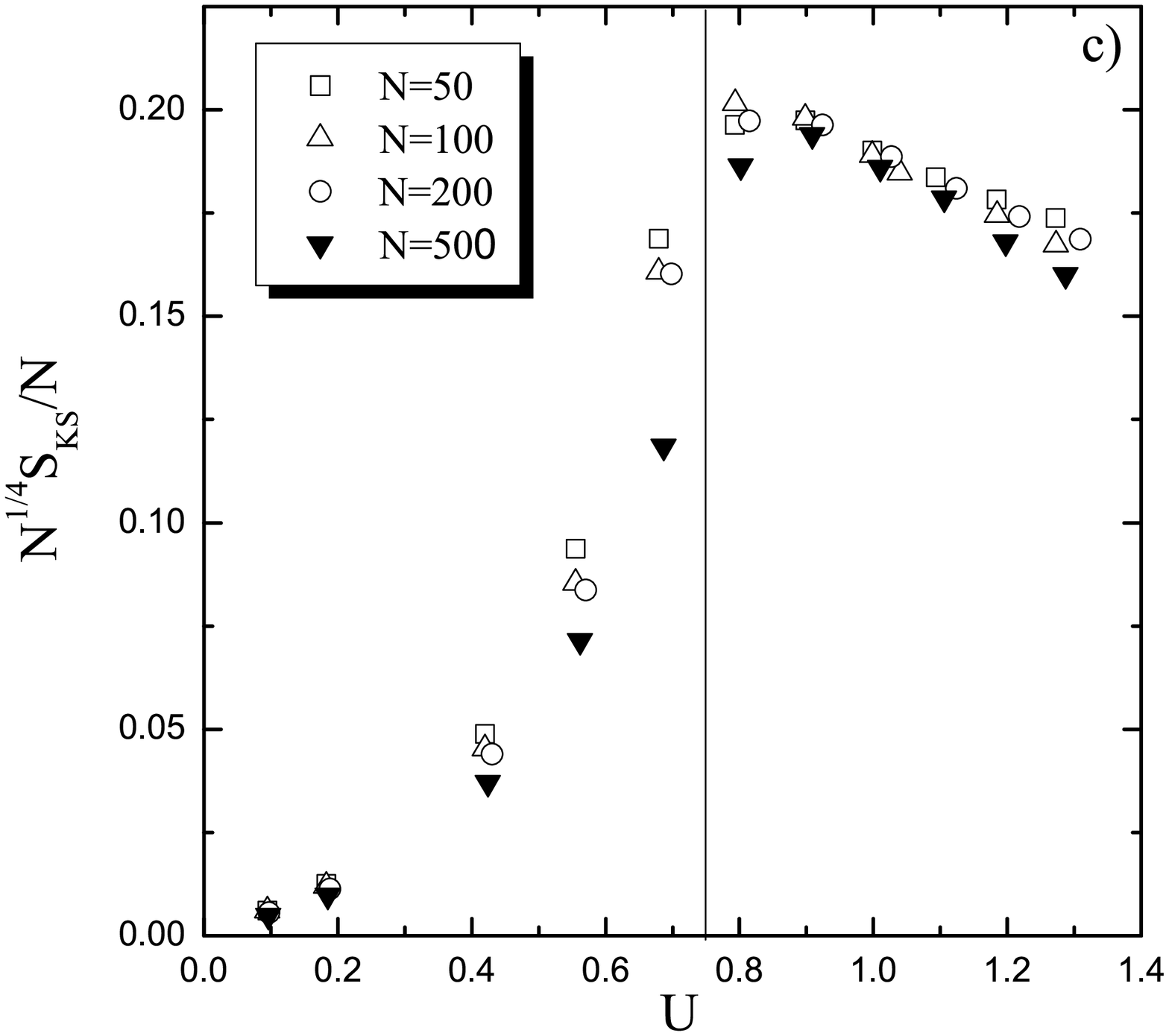}
  \caption{a) KS entropy density vs. energy $U$ for the same data and
  system sizes as in Fig. \ref{LE_ab}.
  b) KS entropy density vs. $U$ multiplied by the factor $N^{1/3}$.
  c) KS entropy density vs. $U$ multiplied by the factor $N^{1/4}$.}
  \label{KS_abc}
\end{figure*}

Since the KS entropy is the sum of all positive Lyapunov exponents,
inheriting some features of the LS, we expect that the scaling should
be revealed more easily.
This is indeed the case, as shown in Fig.~\ref{KS_abc}, where we
plot the KS entropy density $S_{KS}/N$ vs. $U$ for several system sizes (panel a) and the
same quantity multiplied either by $N^{1/3}$ (panel b) or by $N^{1/4}$
(panel c). From panel a) we observe that the KS entropy shows
a general trend to decrease for all energies, as the system size increases.
This is more evident than for the MLE.
Panel b) shows that this size dependence is almost completely eliminated for $U<U_c$,
hence the $N^{-1/3}$ factor captures some essential feature of the scaling
for all energies. However, in the $U>U_c$ range we still observe a
significant size dependence.
This latter is almost completely eliminated in panel c), where
we multiply the KS entropy by $N^{1/4}$, a phenomenological factor
for which we have no theoretical justification. The low energy
KS entropy scaling is however spoiled by this rescaling, showing that
a global scaling, valid for all energies, does not seem to exist.
It is also clear that the KS entropy peak is shifted at energies larger
than $U_c$, at variance with the behavior of the MLE.

%=============================================
%=============================================

The dependence of the MLE on system size will be investigated in more detail
in \citet{Xav_etal:2010}, going to number of particles of the order of
$N=10^6$.

\section{The GALI indices}
\label{GALI_def}

In this Section we will briefly introduce the definition of the
Generalized Alignment Indices (GALI) \citep{GALI:2007} and we
will comment on their physical meaning and their relation with Lyapunov
exponents. We will also present the application of the method to the
HMF model with a limited number of particles, to show its efficiency
for the fast detection of chaotic orbits.

The GALI index of order $p$ (GALI$_p$) is determined through the evolution
of $2 \leq p \leq 2N$ initially linearly independent deviation vectors
$\textbf{w}_i(0), i = 1,2,...,p$. Its calculation is therefore strongly related
to the one of the LS. The evolved deviation vectors $\textbf{w}_i(t)$ are normalized
at given time intervals in order to avoid overflows, but their directions are kept
unchanged. Then, according to \citet{GALI:2007}, GALI$_p$ is defined to be the
volume of the $p$-parallelogram having as edges the $p$ unit
deviation vectors $\hat{\textbf{w}}_i(t)=\textbf{w}_i(t)/\|\textbf{w}_i(t)\|, i = 1,2,...,p$
\begin{equation}
\label{GALI:0}
\text{GALI}_{p}(t)=\parallel \hat{\textbf{w}}_{1}(t)\wedge \hat{\textbf{w}}_{2}(t)
\wedge ... \wedge \hat{\textbf{w}}_{p}(t) \parallel.
\end{equation}
From this definition it is evident that if at least two of the deviation vectors
become linearly dependent during dynamical evolution, the wedge product in Eq.~(\ref{GALI:0})
becomes zero and the GALI$_p$ vanishes.

In the case of a chaotic orbit, all deviation vectors tend to become \textit{linearly dependent},
aligning along the eigenvector corresponding to the maximal Lyapunov exponent and
GALI$_{p}$ tends exponentially to zero following the law \citep{GALI:2007}
\begin{equation}
\label{GALI:1}
\text{GALI}_{p}(t) \sim
e^{-[(\lambda_{1}-\lambda_{2})+(\lambda_{1}-\lambda_{3})+...+(\lambda_{1}-\lambda_{p})]t},
\end{equation}
where the $\lambda_i$ are the Lyapunov exponents (or, better, their finite time approximations).

In the case of regular motion, all deviation vectors move on the
$N$-dimensional tangent space of a torus, on which the motion is
quasiperiodic. Thus, if one starts with $p \leq N$ deviation
vectors, these will remain \textit{linearly independent} on this
tangent space, since there is no particular reason for them to
become aligned. As a consequence, GALI$_{p}$ remains in this case
practically constant for all $p \leq N$. On the other hand, for $p
>N $, GALI$_{p}$ tends to zero, since some deviation vector will
eventually become \textit{linearly dependent}. According to
\citet{ChrisBou,SkoBouAnto}, the GALI indices associated with a
quasiperiodic orbit lying on a $m$-dimensional torus (with $m < N$)
behave as follows
\begin{equation}
\label{GALI:2}
   \text{GALI}_{p}(t) \sim \left\{
                         \begin{array}{ll}
                            \text{constant}, & \hbox{if $2\leq p\leq m$}\\
                           \frac{1}{t^{p-m}}, & \hbox{if $m<p\leq 2N-m$} \\
                           \frac{1}{t^{2(p-N)}}, & \hbox{if $2N-m<p\leq 2N$}.
                         \end{array}
                       \right.
\end{equation}
When $m=N$, GALI$_p$ remains constant for $2 \leq p \leq N$ and decreases to zero
as  $\sim 1/t^{2(p-N)}$ for $N < p \leq 2N$. An efficient way to calculate GALI$_{p}$
is by multiplying the singular values $z_i, i=1,...,p$, computed using the singular value decomposition
procedure of the matrix formed by the deviation vectors $\hat{\textbf{w}}_{i}, i=1,...,p$ \citep{LDI,SkoBouAnto}
\begin{equation}\label{svd_calc}
\text{GALI}_{p}=\prod_{i=1}^{p} z_i.
\end{equation}

The method has been applied successfully to several Hamiltonian
systems like the FPU lattice \citep{SkoBouAnto} and to coupled
symplectic maps \citep{BouManChris}. It has been shown that it efficiently
detects not only regular and chaotic motion but also the dimensionality
of the tori on which the regular trajectory lies.

We begin by analyzing the GALI indices for systems with a few particles and
we show examples of computations performed for the HMF model with $N=2$
(integrable case) and $N=3$.

\begin{figure*}[h!]
\centering
  \includegraphics[width=7.cm]{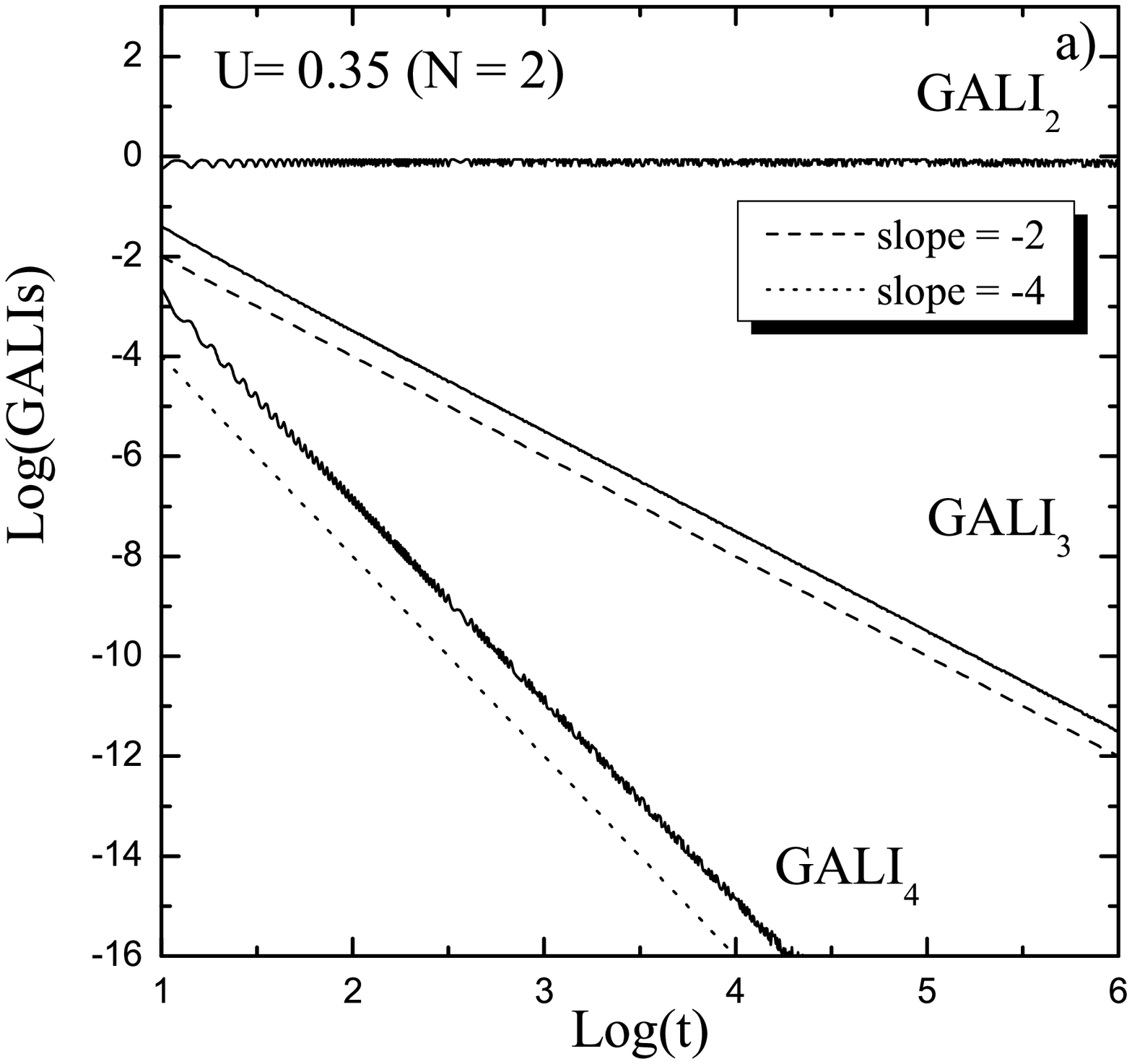}\hspace{-1.75cm}
  \includegraphics[width=7.cm]{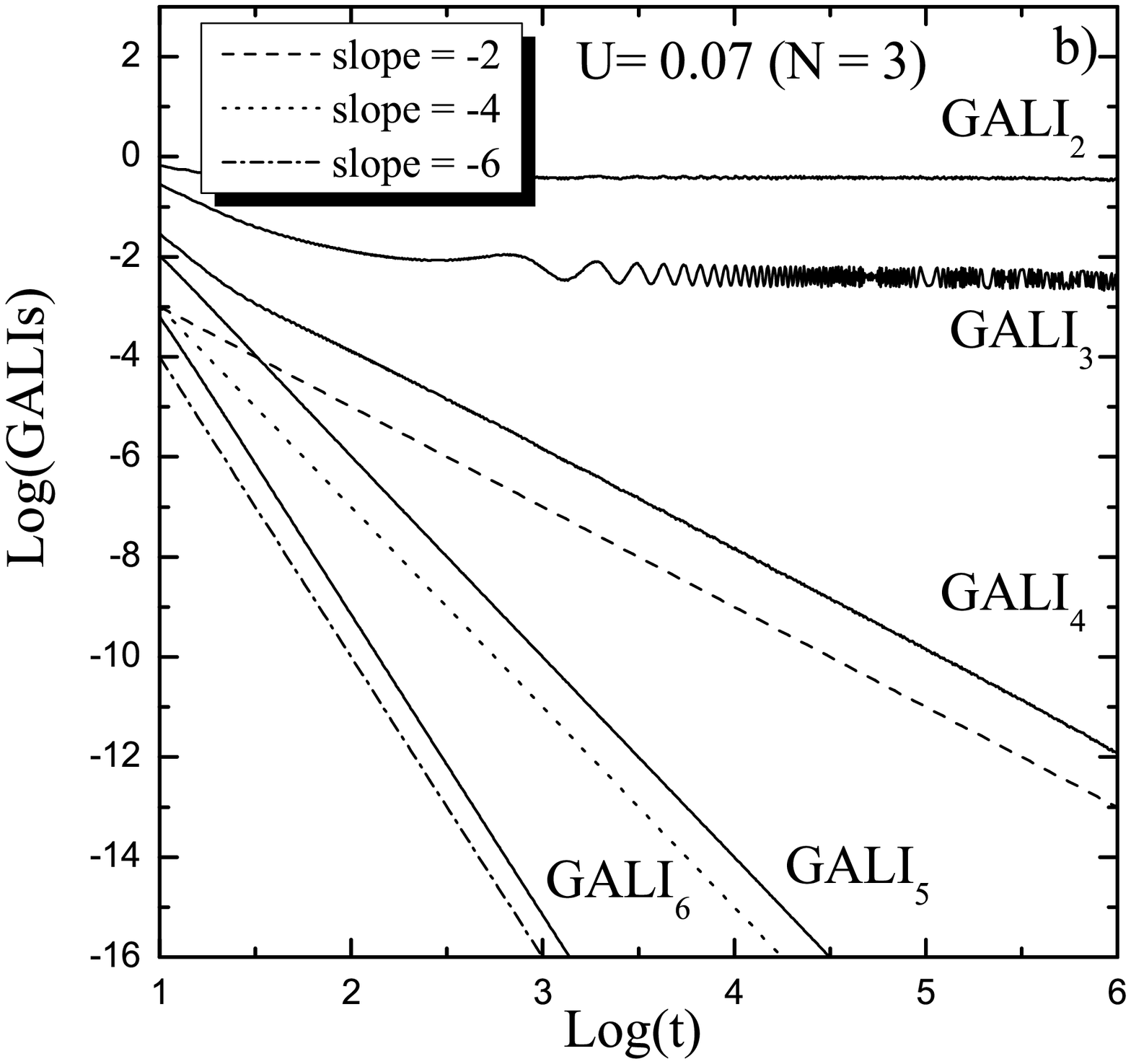}\\
  \includegraphics[width=7.cm]{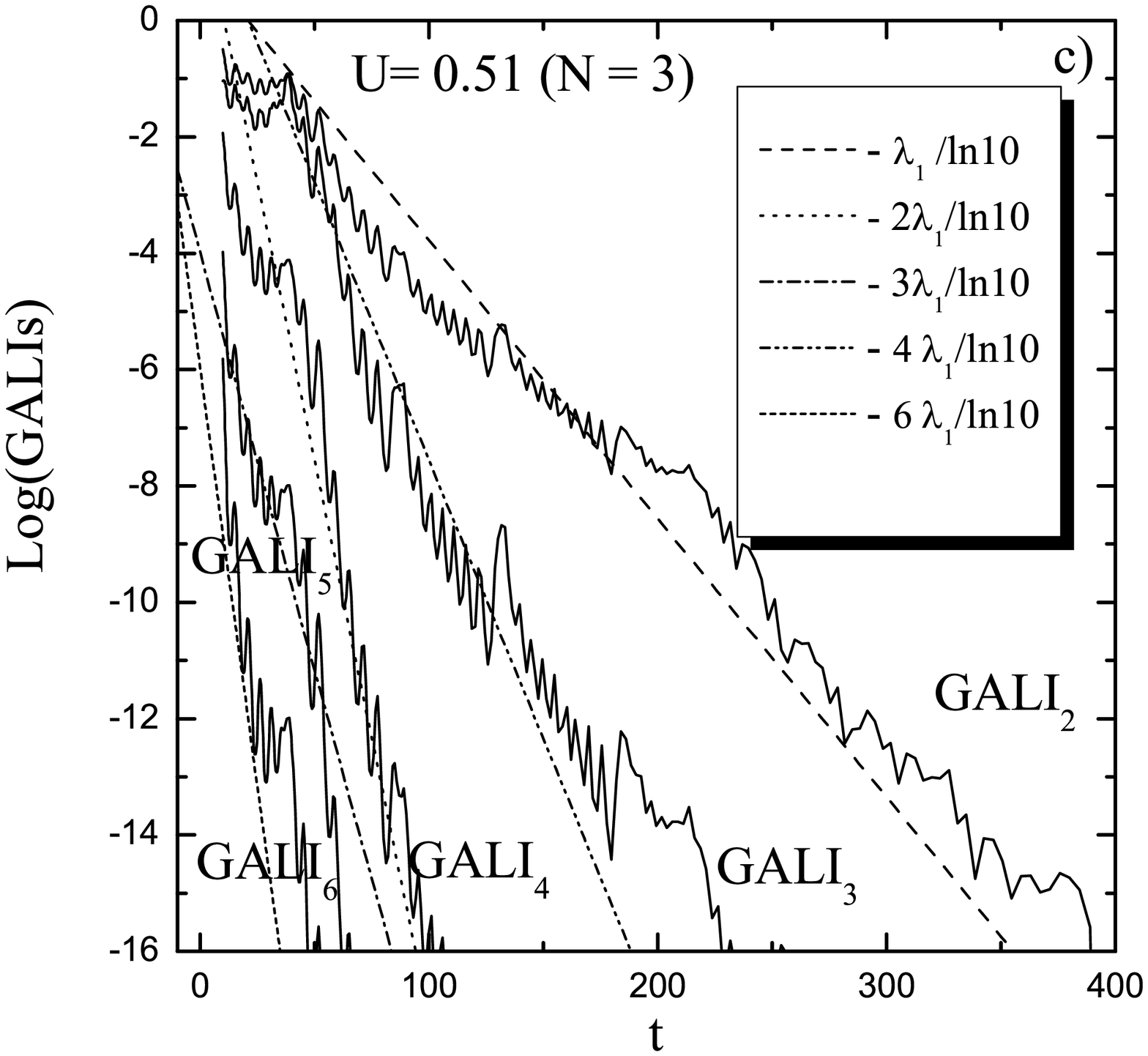}\hspace{-1.75cm}
  \includegraphics[width=7.cm]{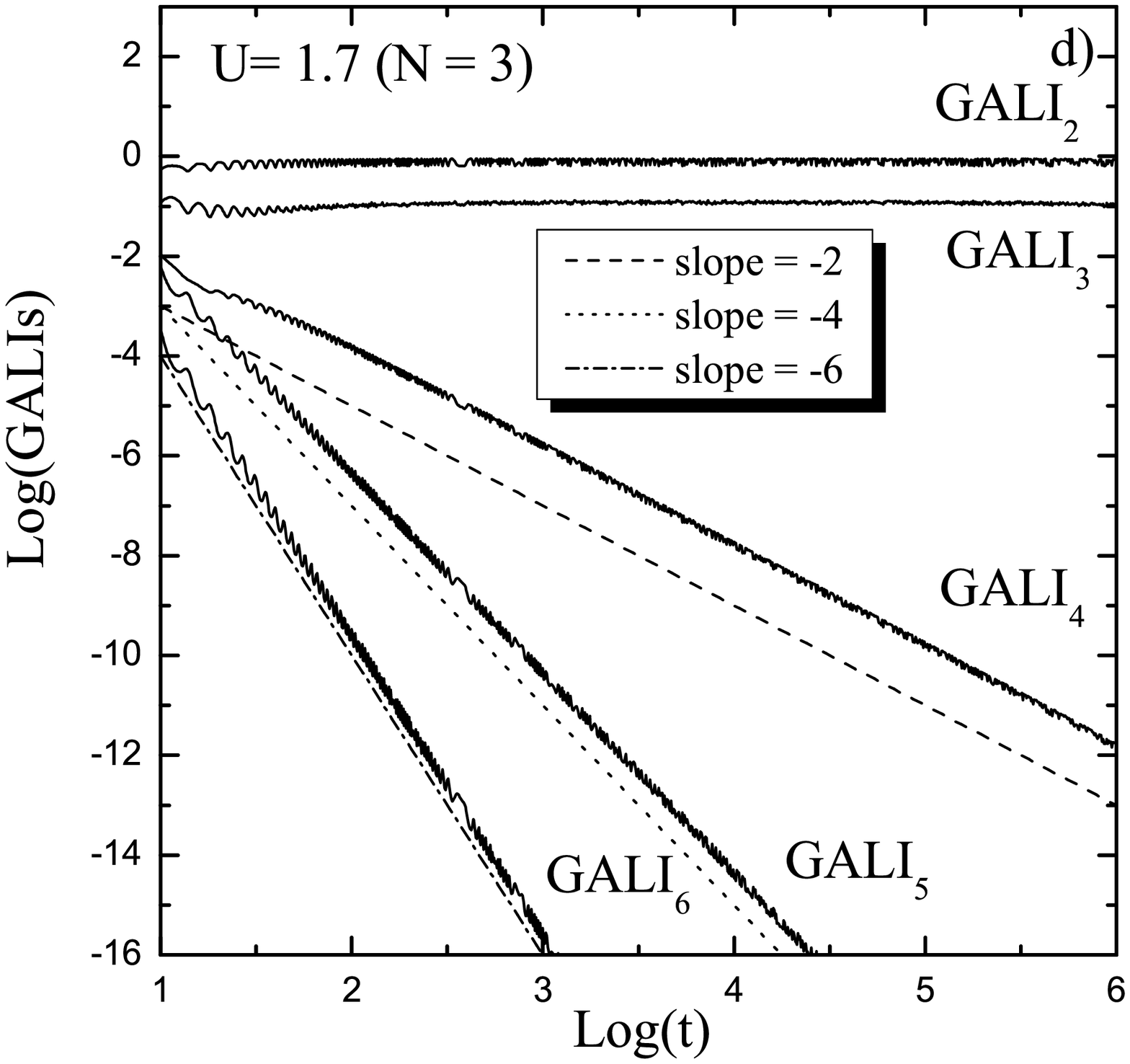}
  \caption{GALI indices vs. time for regular and chaotic orbits.
  a) HMF model with $N=2$ and energy $U=0.35$: integrable system for which GALI$_{3,4}$ decay
  to zero with the power law predicted theoretically.
  b) HMF model with $N=3$ and energy $U=0.07$ for which GALI$_{2,3}$ are constant:
  the motion lies on a 3 dimensional torus.
  c) HMF model with $N=3$ and energy $U=0.51$: chaotic motion is detected, i.e. all GALI's
  decay exponentially to zero. The slopes are given by the Lyapunov exponents.
  d) HMF model with $N=3$ and energy $U=1.7$: the motion is regular and lies again on
  a 3-dimensional torus.}
  \label{GALI:N=2_regular}
\end{figure*}

For $N=2$ we have access to GALI$_{2,3,4}$, since in this case the
dimension of the phase space is $2N=4$. Besides energy, also total
momentum is conserved; hence the system is integrable and the motion
is expected to be regular for all energies. Following the definition
of the GALI indices for regular motion, Eq.~(\ref{GALI:2}), we
should expect
\begin{flalign}
\label{GALI_2Nreg}
\text{GALI}_2(t) &\propto \text{const.}, \quad  \text{GALI}_3(t)\propto \frac{1}{t^2},
\quad \text{GALI}_4(t)\propto \frac{1}{t^4}.
\end{flalign}
Indeed, choosing an initial condition with $U=0.35$ and integrating it
up to $t=10^6$, we see in panel a) of Fig.~\ref{GALI:N=2_regular} that
it produces exactly the predicted time evolution for the GALI$_{2,3,4}$.

For $N=3$ we detect both regular motion (Fig.~\ref{GALI:N=2_regular}b for $U=0.07$ and
Fig.~\ref{GALI:N=2_regular}d for $U=1.7$) and chaotic motion (Fig.~\ref{GALI:N=2_regular}c
for $U=0.51$). In the regular case GALI$_{2,3}$ stay constant, implying a motion that lies on a 3
dimensional torus while GALI$_{4,5,6}$ decay following the power laws
\begin{flalign}
\label{GALI_3Nreg}
\text{GALI}_4(t)\propto \frac{1}{t^2}, \qquad \text{GALI}_5(t)
\propto \frac{1}{t^4}, \qquad  \text{GALI}_6(t)\propto
\frac{1}{t^6}.
\end{flalign}

%=============================================
%=============================================

Chaotic orbits of 3-dimensional Hamiltonian systems generally have
two positive Lyapunov exponents. However, in the HMF model also
total momentum is conserved and, therefore, only one positive
exponent $\lambda_1$ is present in the spectrum. Furthermore, due to
the symmetry of the Lyapunov spectrum $\lambda_6=-\lambda_1$.
Therefore, according to formula~(\ref{GALI:1}), the exponential
decay in time of the GALI indices is controlled only by $\lambda_1$,
as shown in Fig.~\ref{GALI:N=2_regular}c.

The most important advantage of the GALI method relies on the fact
that, for chaotic orbits, the GALI indices decay to zero exponentially
fast. Practically, an orbit can be labeled as chaotic when some
GALI$_p$ (where $p$ is conveniently chosen) is found to be smaller
than a given threshold. This often happens before the MLE stabilizes
on an average positive value, making the GALI$_p$ test more
efficient in determining chaotic properties of an orbit than the
measurement of the MLE. We will give an explicit example of this
property of the GALI indices in the next Section.

%=============================================
\section{Application of the GALI method to the HMF model with large $N$: the transition from weak to strong chaos}
\label{GALI_res}
%=============================================

In this Section we will present the application of the GALI method
to the HMF model, as a tool to reveal the fraction of chaotic orbits
on a constant energy surface with many degrees of freedom. The
dependence of such a fraction on the number of degrees of freedom
has been recently analyzed for fully or partially connected
symplectic maps in \citet{Laveder}, with an application also to the
HMF model.

As we have discussed in the previous Section, the GALI method can be
faster in revealing the chaotic or regular nature of a given orbit
than the calculation of the MLE. In order to perform a systematic
analysis, it is necessary to fix some criteria. The first important
choice is the value of $p$ for the GALI$_p$. In systems with many
degrees of freedom we have access to high values of $p$, but the
calculation of the corresponding GALI$_p$ would be extremely heavy,
since we would have to perform the singular value decomposition for
large matrices. Therefore, we are limited ourselves to small values
of $p$. We have avoided to use GALI$_2$, since it has been shown in
\citet{GALI:2007} that, for some chaotic orbits of many degrees of
freedom systems, the two largest Lyapunov exponents can take very
close values. In such chaotic cases, the GALI$_2$, instead of
decreasing exponentially, stays constant, detecting a ``false"
regular orbit. For the HMF model, in the energy range $0 < U < 0.4$,
we have checked the different time evolution of GALI$_{3,4,5}$ and
we have finally decided to concentrate our attention on the behavior
of GALI$_3$. Fixing the total integration time to $t = 3 \times
10^3$, we have labeled an orbit as regular if GALI$_3$ stays above
$10^{-4}$ over the whole time lapse. We instead label an orbit as
chaotic if GALI$_3$ goes below $10^{-12}$ within the total
integration time. Orbits not satisfying any of these two criteria
are analyzed on a longer time span to assess their chaotic or
regular nature; however, in the simulations we have performed very
few such cases happened. We have decided to fix the number of
initial conditions to $1000$ and we have performed simulations with
increasing number of particles: $N=100,1000,5000$.

\begin{figure*}[h!]
\centering
  \includegraphics[width=8cm]{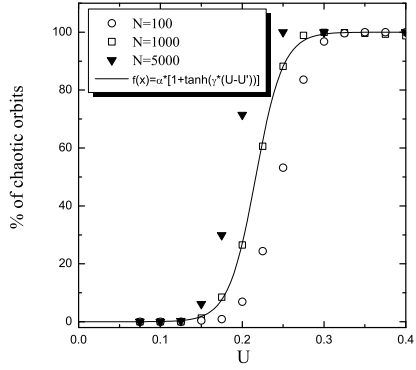}
  \caption{Fraction of chaotic orbits vs. energy $U$ for a different number of particles $N=100,1000,5000$.
  We also show a fit to the points with $N=1000$ particles (full line). The transition energy from weak to strong chaos
  is around $U_t \approx 0.2$.}
  \label{chaos_fraction}
\end{figure*}

In Fig.~\ref{chaos_fraction} we show the fraction of chaotic orbits
in $\%$ as a function of $U$. A sharp increase of this fraction is
observed around the energy $U_t \approx 0.2$. A more quantitative
fit to the data for $N=1000$ with the function
$f(x)=\alpha*[1+\tanh(\gamma*(U-U_t))]$ gives the more precise
estimate $U_t=0.217...$ (the other fitting parameters being
$\alpha=50$ and $\gamma=30$). A significant shift of the transition
towards smaller values of $U$ is observed as $N$ is increased.
However, it is likely that the curve will reach an asymptotic limit
as $N$ is increased (currently not accessible for our computers),
since the shift observed when passing from $N=100$ to $N=1000$ (a
factor of 10) is of the same order as the one found when further
multiplying $N$ by a factor of 5. Moreover, the fraction of chaotic
orbits is below $1 \%$ for $U<0.15$ and around $99 \%$ for $U>0.25$
for all simulated system sizes, showing that a sharp transition from
a {\it weakly chaotic} to a {\it strongly chaotic} phase-space takes
place in the energy range $[0.15,0.25]$.

%=============================================
%=============================================

It has been suggested by \citet{Ant_Ruf:1995} that, around energy $0.2$,
the phase $\phi$ of the order parameter $\mathbf{M}$ of the HMF model
begins to be strongly time dependent, while for smaller energies it is
essentially time independent. This determines the untrapping of some
particle trajectories from the nonlinear resonance in the pendulum phase-space
associated to the model. Since we here find a significant growth
of the fraction of chaotic orbits just in this energy range, we wanted to
verify the correspondence between the two phenomena.

\begin{figure*}[h!]
\centering
  \includegraphics[width=6.cm]{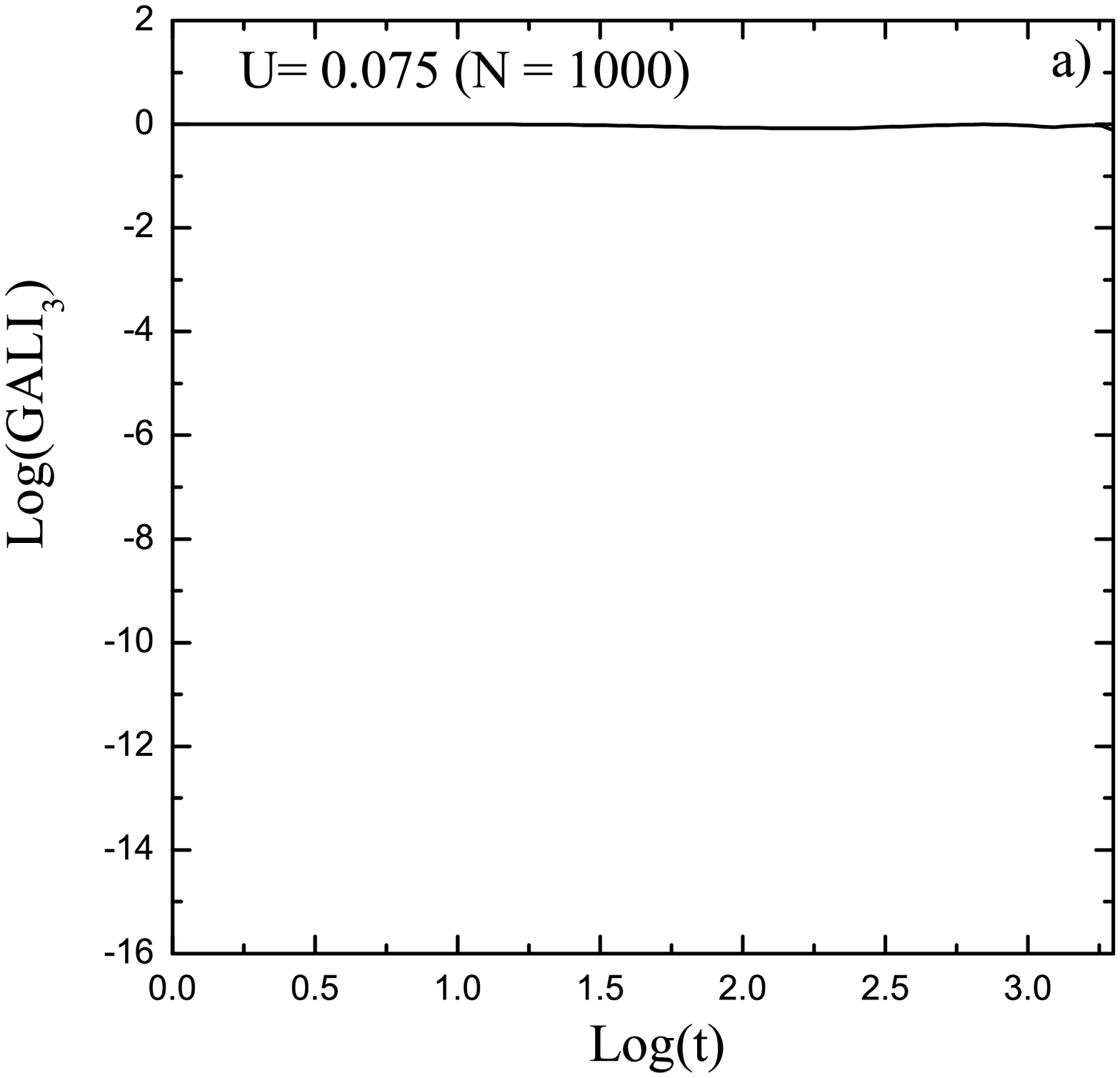}\hspace{-1.75cm}
  \includegraphics[width=6.cm]{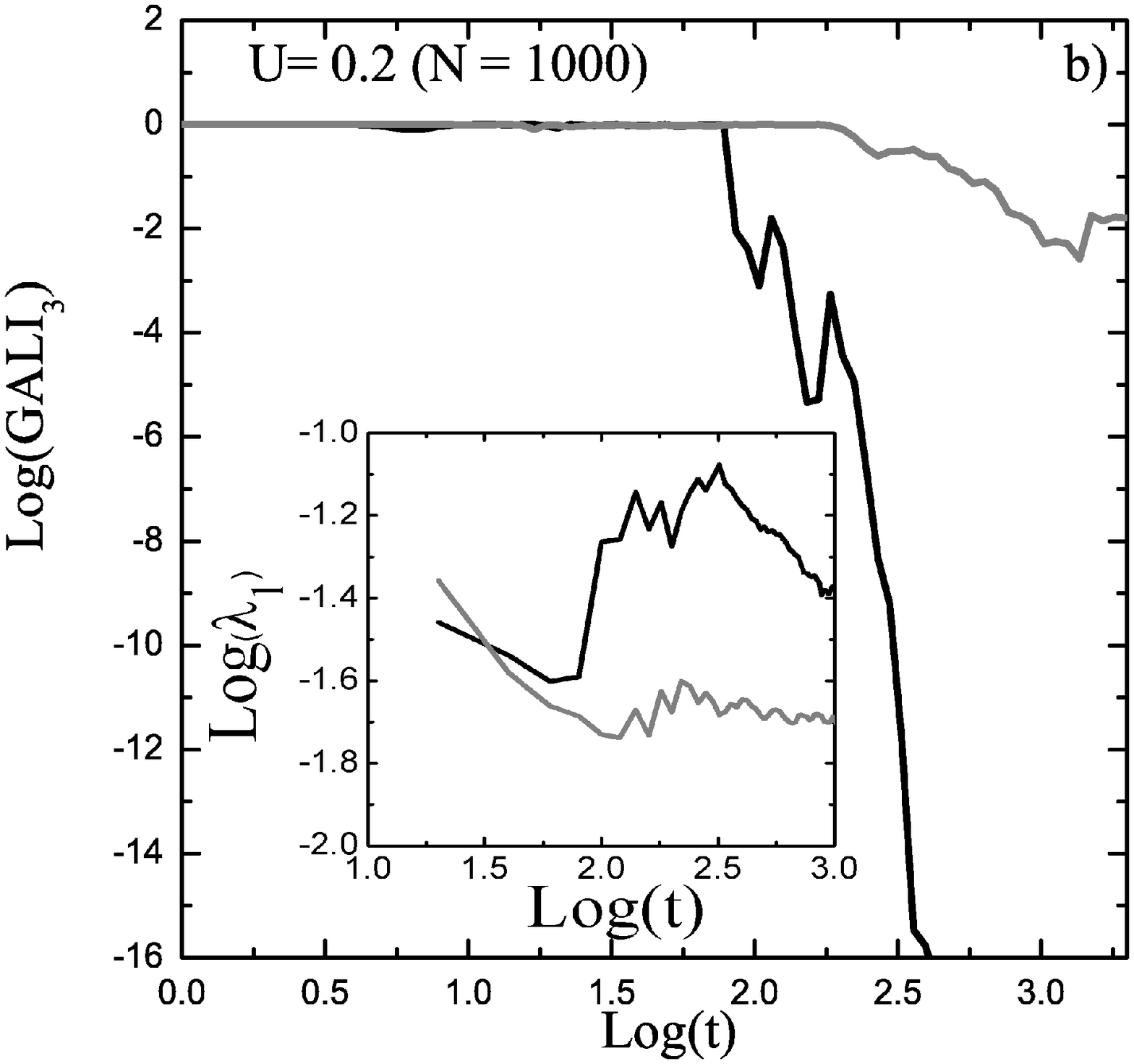}\hspace{-1.75cm}
  \includegraphics[width=6.cm]{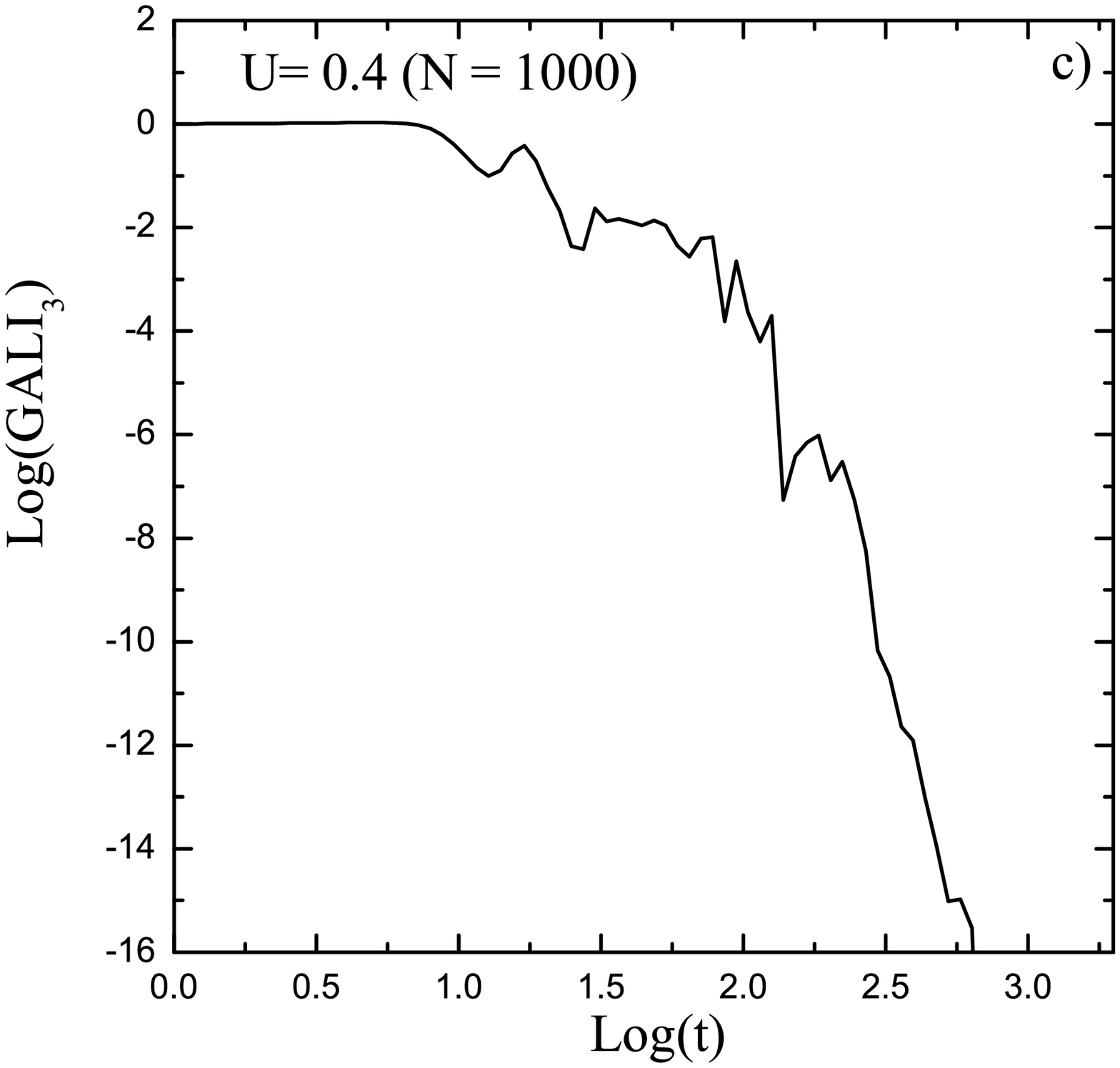}\\
  \includegraphics[width=6.cm]{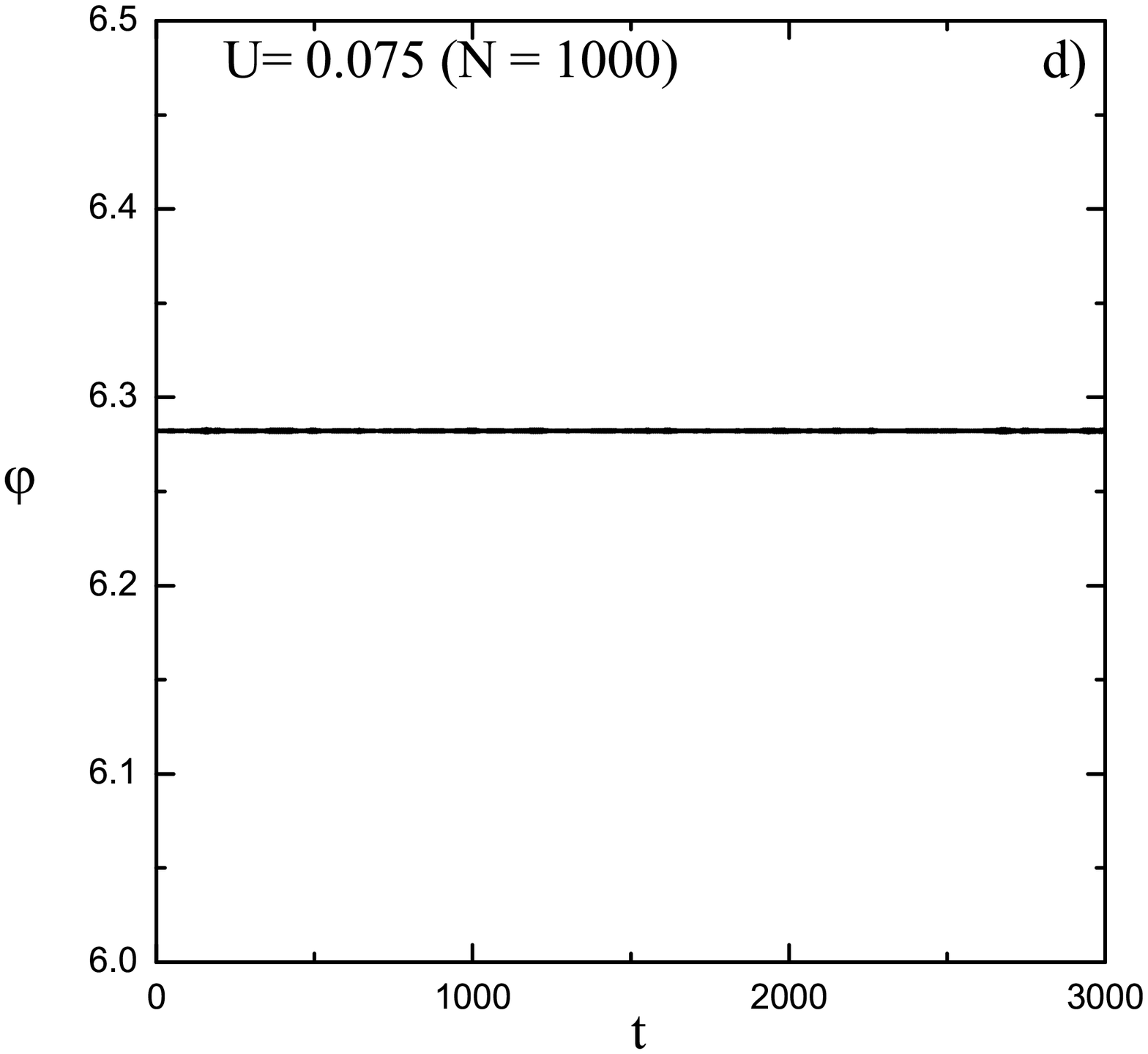}\hspace{-1.75cm}
  \includegraphics[width=6.cm]{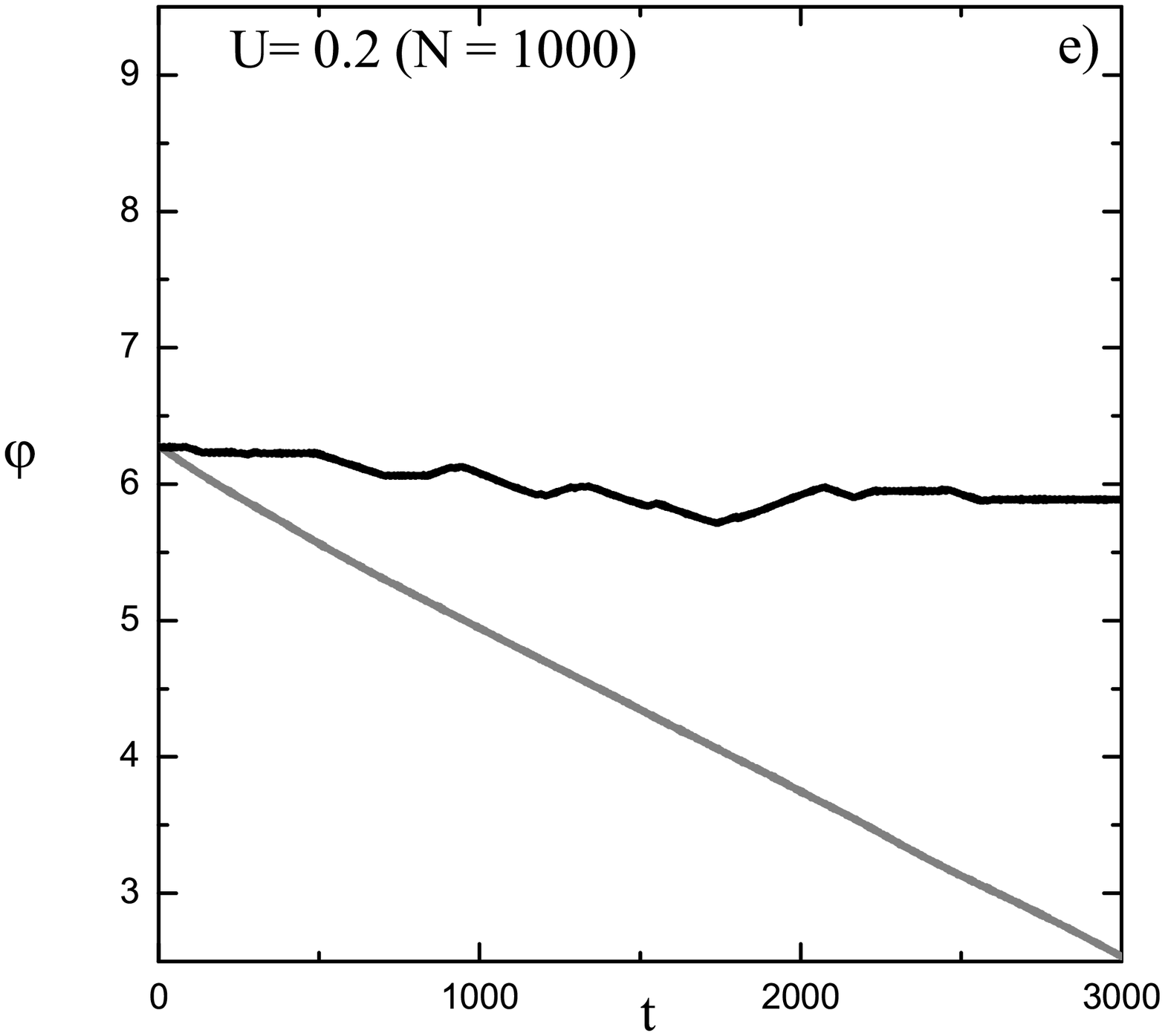}\hspace{-1.75cm}
  \includegraphics[width=6.cm]{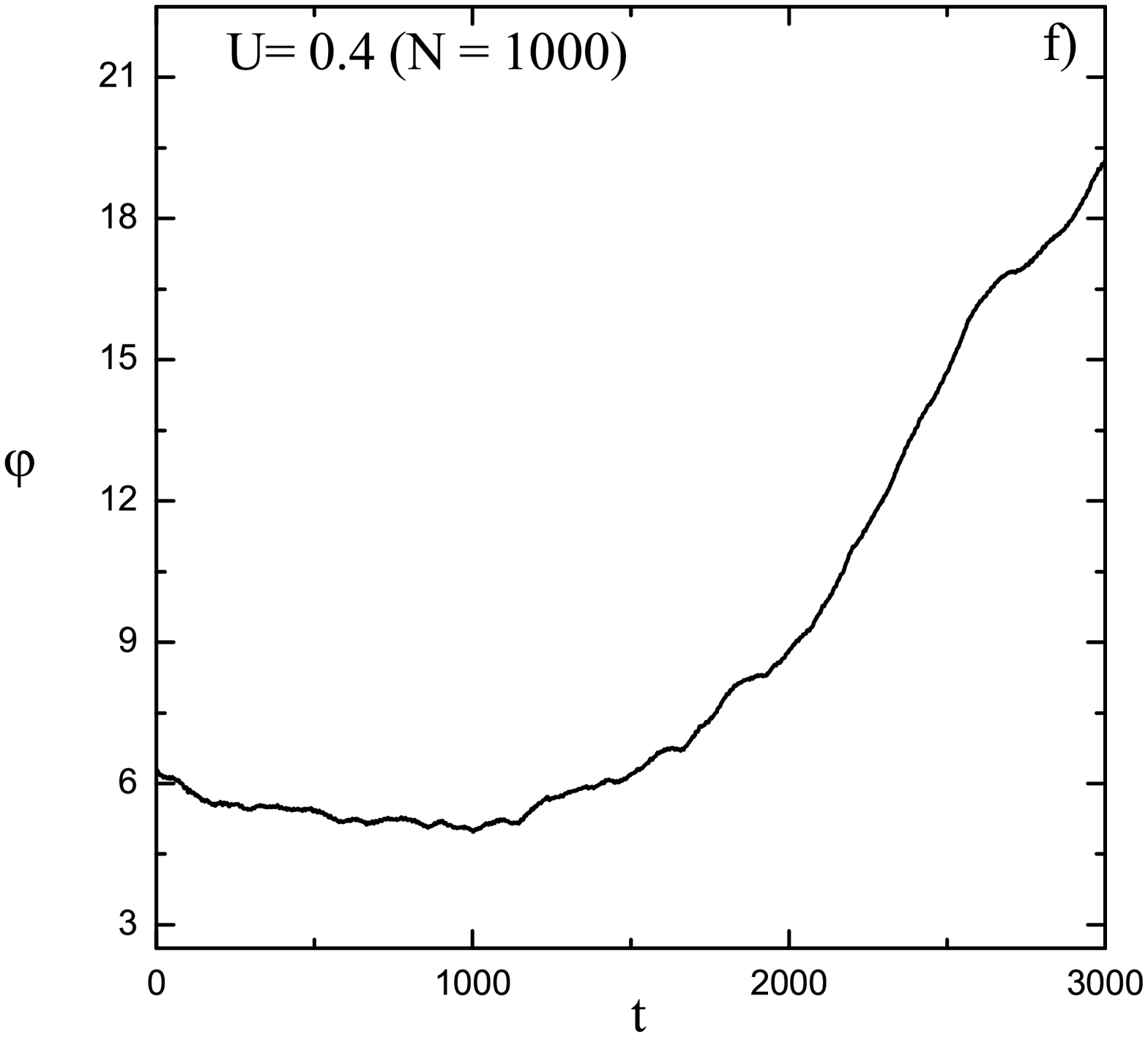}
  \caption{GALI$_3$ vs. time (upper row) and phase $\phi$ of the order parameter
  $\mathbf{M}$ (see Eq.~\ref{magn}) vs. time (lower row) at increasing energies for $N=1000$.
  a) GALI$_3$ remains constant in time for a regular orbit at $U=0.075$.
  b) GALI$_3$ for a chaotic orbit (thick full line) and for a less chaotic one, classified
  as regular according to our criterion (thin full line), at $U=0.2$.  In the inset we show the
  corresponding time evolution of the MLE.
  c) GALI$_3$ decays exponentially for a typical chaotic orbit at $U=0.4$.
  d) Phase $\phi$ of the order parameter vs. time for $U=0.075$ computed for the orbit of panel a).
  e) Phase $\phi$ of the order parameter vs. time for $U=0.2$ computed for the two orbits of panel b).
  f) Phase $\phi$ of the order parameter vs. time for $U=0.4$ computed for the chaotic orbit of panel c).}
   \label{GALI3_phase}
\end{figure*}

In Fig.~\ref{GALI3_phase} (upper row) we plot the time evolution of
GALI$_3$ for three different energies: $U=0.075, 0.2, 0.4$. At low
energy (panel a) the GALI$_3$ remains constant, revealing a regular
orbit, while at high energy (panel c) it decreases exponentially
fast, detecting a chaotic orbit. In the energy region around $U_t$,
part of the orbits are chaotic and part are regular. We here show
(panel b), two orbits that, according to our criteria, are labeled
as chaotic (thick full line) and regular (thin full line). In the
inset of panel b) we plot the time evolution of the MLE for these
two orbits: the MLE of the chaotic orbit is larger that the one of
the ``regular" orbit, but on this time scale, they are still far
from convergence. This proves that the GALI$_3$ is able to detect
the regular and chaotic nature of the orbit faster than the MLE.

Phase motion is absent for the regular orbit at low energy, see
Fig.~\ref{GALI3_phase}d, while it's clearly present for the chaotic
orbit at the highest energy in Fig.~\ref{GALI3_phase}f. At the
energy $U=0.2$ (Fig.~\ref{GALI3_phase}e) both the chaotic and the
regular orbit display phase motion but, surprisingly, the regular
orbit shows a ``ballistic" motion of the phase. It's unclear why
this happens; a possible explanation is that, for this orbit,
untrapped particles remain out of resonance for a longer time. This
would imply less chaoticity because some particles remain far from
the most chaotic part of the phase-space, which is around the
chaotic layer of the separatrix.

%=============================================
%=============================================

Although the understanding of the transition from {\it weak} to {\it
strong} chaos appearing around the energy $U_t$ is far from being
satisfactory, its numerical evidence is clear. Moreover, this
threshold energy separates an energy region, $0<U<U_t$, where the
$N^{-1/3}$ scaling of the MLE and of the LS is clearly found from a
higher energy region, $U_t<U<U_c$, where this scaling is not found
with the same evidence. Indeed, in this intermediate energy range,
the MLE hardly converges to zero as the system size is increased, as
shown in Fig.~\ref{LE_ab}a.

\section{Conclusions}
\label{concl}

Although the chaotic properties of the Hamiltonian Mean Field (HMF) model
have been studied in several papers, a general understanding of the
behavior of the Lyapunov exponents as the system size is increased and
for different energies is still lacking.

We have here shown that, rather than concentrating on the Maximal
Lyapunov Exponent (MLE), it is sometimes preferable to determine the
full Lyapunov Spectrum (LS). This is certainly more challenging from
the numerical point of view, but can be rewarding and reserve some pleasant surprises.
We here show, for instance, that the convergence of the LS to its
asymptotic, large $N$, shape is more rapid than the one of the MLE.

Moreover, by the study of the LS, the general scaling law with
$N^{-1/3}$, originally proposed by \citet{Latoraetal:1998}, is here
found for moderate system sizes. Although this law was originally
proposed for the high energy phase of the HMF, above the critical
energy $U_c=3/4$, we show in this paper that it is also valid in a
wide range of low energies, below $U_t \approx 0.2$.

With the aim of investigating the physical meaning of this threshold
energy $U_t$, we have used the newly proposed GALI method by
\citet{GALI:2007} in order to determine the fraction of chaotic
orbits in the phase-space of the HMF model. The method reveals that
this fraction sharply changes from few percents to almost $100 \%$
in an energy range around $U_t$, showing a transition from {\it
weak} to {\it strong} chaos in this energy region. It is for
energies $0<U<U_t$ that the $N^{-1/3}$ law is best verified. On the
contrary, when $U_t<U<U_c$, the scaling law is numerically unclear
and the MLE shows a slow decrease to zero as the system size
increases, which could even lead to suspect that the MLE remains
positive in the $N \to \infty$ limit.

Above $U_c$ one observes a general trend to decrease to zero of the full LS.
The scaling is precocious if one studies full the LS and is again well fitted by the
$N^{-1/3}$ law. The MLE also decreases to zero, but with an intermediate
scaling with $N^{-1/4}$. It is only for much larger system sizes, of the order
of $N=10^6$, that one finally finds the $N^{-1/3}$ law also for the MLE.

On the theoretical side, the only existing approach valid in
principle for all energies \citep{Firpo:1998} predicts a strictly positive
MLE for $0<U<U_c$, while the MLE should vanish as $N^{-1/3}$ for
$U>U_c$. This is not what we find for $0<U<U_t$, demanding for an
improvement of this theoretical approach. In the high energy region
$U>U_c$, a random matrix approximation
\citep{Latoraetal:1998,Firpo:2001,Anteneodo1:2001,Anteneodo2:2002,Anteneodo3:2003}
predicts the $N^{-1/3}$ scaling of the MLE, which has been checked
numerically by
\citet{Latoraetal:1998,Ante_Tsa:1998,Anteneodo3:2003}. It is hard to
believe that such an approach could be used for the energy range
$0<U<U_t$, where the phase-space is mostly occupied by regular
orbits, as the GALI method allowed us to assess. The understanding
of why the $N^{-1/3}$ scaling is valid in this energy region remains
a challenge for future research.

%=============================================
\section{Acknowledgements}
We would like to thank H. Chat\'e, D.~Fanelli, F. Ginelli, X.~Leoncini, R. Paskauskas,
A.~Politi and Ch.~Skokos for fruitful comments and discussions. We acknowledge financial
support of the COFIN-PRIN program ``Statistical physics of strongly correlated
systems at and out of equilibrium" of the Italian MIUR.

%=============================================

\end{document}